\documentclass[aps,prb,twocolumn,showpacs,floatfix]{revtex4}
\usepackage{latexsym}
\usepackage{graphicx}
\newcommand{\figwidth}{2.4 in}
\newcommand{\figw}{2.2 in}
\usepackage{subfigure}    
\usepackage{epsfig}
\begin{document}
\title{Convergence of quasiparticle self-consistent  GW calculations 
of transition metal monoxides}
\author{Suvadip Das$^{1}$}
\author{John E. Coulter$^{1}$}
\author{ Efstratios Manousakis$^{1,2}$ }
\affiliation{
$^{(1)}$ Department  of  Physics and National High Magnetic Field Laboratory,
  Florida  State  University,  Tallahassee,  FL  32306-4350,  USA\\
$^{(2)}$Department   of    Physics,   University    of   Athens,
  Panepistimioupolis, Zografos, 157 84 Athens, Greece
}
\date{\today}
\begin{abstract}
Finding an accurate {\it ab initio} approach for calculating
the electronic properties of transition metal 
oxides has been a problem for several decades.
In this paper, we investigate the electronic 
structure of the transition metal monoxides MnO, CoO, and NiO in their
undistorted rock-salt structure within a 
fully iterated quasiparticle self-consistent  GW (QPscGW) scheme. We study 
the convergence of the
QPscGW method, i.e., how  the quasiparticle energy eigenvalues 
and wavefunctions 
converge as  a function of the QPscGW iterations, and we compare the converged 
outputs obtained from different starting wavefunctions. 
We find that  the convergence is slow and that a one-shot G$_0$W$_0$ 
calculation does not significantly improve the initial eigenvalues and states. 
It is important to notice that in some cases the ``path'' to convergence may 
go through energy band reordering which cannot be captured by the simple 
initial unperturbed Hamiltonian.
When we reach a fully iterated solution, the converged density 
of states, band-gaps and magnetic moments of these 
oxides are found to be only weakly dependent on the choice of the starting  
wavefunctions and in reasonably good agreement with the experiment. 
Finally, this approach provides a clear picture of the interplay between
the various orbitals  near the Fermi level of these simple transition 
metal monoxides.  The results of these accurate {\it ab initio}
calculations can provide input for models aiming at
describing the low energy physics in these materials.

\end{abstract}
\maketitle

\section{Introduction}
Transition metal oxides (TMO) form a very interesting class of materials 
exhibiting a rich variety of physical properties resulting from the 
interplay of spin, orbital, charge and lattice dynamics. They form the 
building block of many materials showing complex behavior including 
superconductivity, colossal magnetoresistance and multiferroic behavior. 
Among the 
TMOs, MnO, CoO, and NiO,  have been extensively studied, because they 
are the simplest TMOs with incomplete $d$ shells.
Furthermore, they have been used as a playground for applying new 
\emph{ab initio}
methodology  because for some of them, simple density functional 
theory (DFT) within the local density approximation (LDA) does not yield 
the correct character of their ground state. 
While originally they were 
thought to be Mott insulators\cite{Mott}, later 
studies\cite{Terakura,Oguchi} indicate that these materials might be
charge transfer insulators.\cite{Zaanen} 
These oxides are found in 
the rock-salt structure in the paramagnetic phase and undergo 
antiferromagnetic ordering below their Neel temperature along with 
structural distortions. 

Furthermore, understanding these materials is crucial because 
various applications are being
explored using TMOs\cite{VO2-Device}, including utilizing 
their optoelectronic properties.\cite{Mottsolar,Miller12} 

The Kohn-Sham DFT within the LDA has provided a very successful {\it ab initio}
framework to successfully tackle the problem of the electronic structure of
materials. However, shortly after the discovery of the copper-oxide 
superconductors,
certain weaknesses of the method were exposed, as it failed to yield the
fact that the parent compound La$_2$CuO$_4$  is an antiferromagnetic insulator.\cite{Warren}
Furthermore, this particular approximation also
fails to yield the insulating character of simple $d$ electron transition
metal monoxides, such as CoO\cite{Oguchi}, which is the case of our interest in this 
paper.  This difficult period for the DFT/LDA method was partially ended
in the early and mid 90s when an orbital dependent Hubbard type U was 
incorporated in the exchange correlation functional of the localized 
$d$ electrons in a mean field fashion within the (LDA) + U 
method\cite{Anisimov,Liechtenstein}, while the itinerant electrons are 
still described at the LDA level. Although the LDA + U method has been 
successful in treating  localized electron systems, the results are 
strongly dependent on the choice of the parameter U.

Another approach to the problem is the so called GW approximation of 
Hedin \cite{Hedin} which yields an approximation to the single-particle 
Green's function and 
takes many-body effects into account in the electron-electron interaction. 
This many-body perturbation technique not only supports a quasiparticle 
picture, but also accounts for the dynamical screening of the 
electrons. 

This is achieved by approximating the single-particle self-energy
as $\Sigma=iG W$, in terms of the single-particle Green's 
function $G$ and the dynamically screened Coulomb interaction $W$, which 
is obtained using the inverse of the frequency-dependent dielectric matrix.
In spite of the neglect of vertex corrections in the self energy, 
which gives rise to higher order correction terms and overestimation of 
the band gaps due to underestimated dielectric constants\cite{Shishkin3}, 
the GW calculations 
give very good agreement between calculated and 
measured band-gaps (as well as other single-particle properties) for 
$sp$ semiconductors.\cite{Schilfgaarde,Hybertsen} 

In the case of systems where localized $d$ (or $f$) states are present near
the Fermi energy, as in this work, 
a large number of plane wave states are required to accurately represent 
these localized states.
In order to tackle this problem, Gaussian orbitals or localized 
basis sets have been used within the linear muffin-tin-orbital (LMTO)
method\cite{LMTO-origin}.
Such an approach has been repeatedly applied 
successfully to NiO\cite{Aryasetiawan,SMA-NiO}, which is a subject of the 
present paper. There are other methods to circumvent the problem, such as 
full-potential LMTO or full-potential linear augmented plane-wave 
methods\cite{Andersen-LAPW}, 
which use a localized basis
set in the muffin-tin region and plane waves in the interstitial region.
The projector augmented-wave (PAW) 
method\cite{Blochl} as implemented in the VASP 
code\cite{Shishkin3,Fuchs,Shishkin2,Shishkin1} works quite efficiently
in conjunction with the GW method. 

There is a significant effort 
to apply the GW approach to the  TMOs. 
There are calculations performed for NiO using the 
spin-polarized GW 
approximation in the LMTO basis set \cite{Aryasetiawan-spectral,Aryasetiawan}, 
and a plane-wave basis set with ab initio 
pseudopotentials\cite{Louie_SG}.
The so-called ``model GW'' has been used to investigate 
the single-particle properties of  MnO and NiO\cite{Massidda} and  of MnO, 
FeO, 
CoO and NiO\cite{Asahi}.  All-electron self-consistent quasiparticle 
GW calculations were performed on MnO and 
NiO\cite{Faleev, Schilfgaarde}. 
A GW calculation\cite{Kobayashi} of the single-particle properties 
starting from LDA+U wavefunctions has been performed 
on NiO and MnO and a good agreement with experiment was achieved. 
With a somewhat similar approach, starting from a LDA+U calculation which
was followed by G$_0$W$_0$ and GW$_0$ calculations (in which G and W, or 
just W, is calculated only at the 0th order), the band gap and 
self-energy of these oxides were also investigated\cite{Rinke}. 

It has been suggested that wave functions obtained within hybrid functional
calculations (HSE)\cite{HSE} may provide a good starting point for 
GW calculations\cite{Rodl-G0W0+HSE,Bechstedt-New,Fuchs08}.
The band structures of MnO, 
FeO, 
CoO and NiO have been studied within 
the HSE+G$_0$W$_0$ approach and reasonable agreement with experimental 
band gaps was found\cite{Rodl-G0W0+HSE}. 
However, a recent publication\cite{LanyOxides} shows that the choice of 
HSE functional as the starting wavefunction produces an incorrect band 
ordering in Cu$_2$O, which can be resolved by means of a self-consistent 
GW calculation. Another comprehensive work on hybrid functionals 
found that the basic HSE06 functional was insufficient for many materials
\cite{Marques1}.

Because HSE might be considered a good starting point,
one might think that it is sufficient to carry out only a few additional
GW iterations.
However, it was found\cite{Coulter-scGW} that in order to obtain accurate 
wavefunctions, especially for states near the Fermi energy, one might need 
to carry out many GW iterations and, thus, this approach is both
computationally costly and \emph{not} parameter free. 
A recent fully iterated quasiparticle self-consistent GW calculation
gives the correct band gap and quasiparticle wavefunctions, 
\emph{independent} of 
the starting LDA and HSE wavefunctions for the transition metal 
oxide VO$_2$\cite{Coulter-scGW}.
Earlier works using a similar method 
demonstrated that self-consistency is very important in 
VO$_2$\cite{SMA-GW,SMA-VO2-GW}. Additionally, quasiparticle lifetime
calculations have been successfully performed in this material by
QPscGW, though very high
precision is required for such work\cite{Coulter-IIR}.

The method of LDA+DMFT has been consistently
useful in the case of strongly correlated TMO's. It has been fruitful 
in these
materials for many types of study; from accurate electronic and
magnetic structure\cite{NiO-DMFT1,NiO-DMFT2,All4-DMFT1,LDA'+DMFT} to 
exploring metal-insulator transitions\cite{MnO-DMFT1,MnO-DMFT2,CoO-DMFT1}
 and the effect of doping\cite{NiO-DMFT3}.
As we will
discuss in the conclusions section of this paper, these DMFT
results associated with character the bands and their ordering near the 
Fermi level are in agreement with our QPscGW results for the materials
we consider in this paper.

In this work we focus our effort to answer
a few pressing questions regarding the QPscGW approach to be 
adopted for later studies of other TMOs:
Are too many QPscGW steps required for convergence to make the approach 
practical?
How does the convergence rate depend on the initial quasiparticle 
wavefunctions, i.e., wavefunctions and energies 
obtained from a GGA or GGA+U or HSE calculation? 
How profitable is it to carry out a single shot G$_0$W$_0$ calculation on 
top of GGA, or GGA+U or HSE for this class of materials?
How strongly do the converged solutions depend on these 
initial choices of quasiparticle wavefunctions? Can we approximate the
results of the fully converged QPscGW approach with a GGA+U calculation
in an appropriate regime of U?

To answer these questions we focus our effort on MnO, NiO and CoO for the
following reasons.
In the case of MnO and NiO the spin-polarized GGA (sGGA) calculation yields a 
band-gap. This allows us to start with a 
QPscGW\cite{Shishkin3,Fuchs,Shishkin2,Shishkin1} based on wavefunctions and 
quasiparticle energies obtained by means of a spin-polarized GGA 
calculation. Similar QPscGW calculations have given good results on similar
materials\cite{Svane2010,Coulter-scGW}, including strongly 
correlated $f$ electron systems\cite{Chantis2007,Chantis2008}. 

However, there are many insulating materials, especially in the class of 
TMOs, for which a simple sGGA calculation fails to show a gap. 
CoO is such an example, where sGGA yields no gap.
In order to accelerate the rate of convergence 
of the GW method for CoO, we start the QPscGW iterations 
using the wavefunctions obtained from a GGA + U calculation.
If such calculations are fully iterated, the results should be
only very weakly dependent on the initial value of U. 
In this paper, we have investigated the effect of such starting 
wavefunctions on the self consistent GW calculations. We explore the 
convergence with different starting values of U for CoO. For the case of 
MnO and NiO we explore the convergence starting from wavefunctions and 
energies obtained from GGA and HSE calculations.

The paper is organized as follows: 
In Sec.~\ref{methods} we discuss the computational approach and 
the details of the scheme adopted here.
In Sec.~\ref{convergence} we discuss  details of the convergence of
the QPscGW approach for all three TMOs chosen for the present study.  
In Sec.~\ref{results}
we present our converged results for bands, gaps, magnetic moment and the
density of states for MnO, CoO, and NiO, and we compare 
them with the experimental results.
In Sec.~\ref{conclusions} we discuss the main conclusions of the 
present study.

\section{Computational Approach}
\label{methods}
\subsection{The self-consistent GW method}

The QPscGW calculations are performed using a variant of the method 
originally suggested by van Schilfgaarde {\it et al.}\cite{Schilfgaarde} 
as implemented in the VASP code and as outlined in 
Ref.~\onlinecite{Shishkin3}. 
Notice that the formalism presented in 
Ref.~\onlinecite{Shishkin3}, with the interpretation that the state $n$ is 
an abbreviation which includes all orbital and spin
degrees of freedom, allows us to carry out a QPscGW in which ``up'' and 
``down'' spin states are handled as in standard spin-polarized DFT.
In the current version of the VASP code, 
the particular vertex correction contribution to the
frequency dependent dielectric matrix described in 
Ref.~\onlinecite{Shishkin3} has not been implemented for the spin polarized 
case.

In order to perform our QPscGW calculations, we choose a semi-local exchange 
correlation potential within the sGGA (for MnO and NiO)
or the GGA+U (for CoO) approximation 
as the starting point  and we solve the GW equations iteratively as 
discussed in Ref.~\onlinecite{Shishkin3}. 
The exchange correlation potential is updated at every iteration $i$ by 
linearizing the self-energy $\Sigma^{(i-1)}(\epsilon)$ obtained in the  
iteration $i-1$
near the known quasiparticle energy eigenvalue $E^{(i-1)}_n$ 
obtained in the previous  step. 
\begin{equation}
\textbf{H}^{(i)}\mid \psi^{(i)}_{n}\rangle = E^{(i)}_n \textbf{S}^{(i)} | \psi^{(i)}_{n}\rangle,
\end{equation}
where the Hamiltonian and overlap matrices are given by:
\begin{eqnarray}
\textbf{H}^{(i)}&=&T\; +\; V_{ext}\; +\; V_{H}\; +\; V_{xc}^{(i)}, 
\label{hamiltonian}\\ 
V_{xc}^{(i)} &=& [\Sigma^{(i-1)}(\epsilon)\; -\; \epsilon \frac{\partial \Sigma^{(i-1)}(\epsilon) }{\partial {\epsilon}}]|_{\epsilon=E_{n}^{(i-1)}}, \label{exc-corr} \\
\textbf{S}^{(i)} &=& [1 -\; \frac{\partial \Sigma^{(i-1)}(\epsilon)}
{\partial \epsilon}|_{\epsilon=E_{n}^{(i-1)}} ].
\end{eqnarray}


There is no unique method to map this problem onto a corresponding 
Hermitian eigenvalue problem; the
Hamiltonian operator $\textbf{H}^{(i)}$ and the overlap operator 
$\textbf{S}^{(i)}$ can be expressed in a suitable basis set
$|\phi_n \rangle$ (e.g., the DFT wave functions), and we take
the Hermitian part of the self-energy and overlap matrix
in this basis\cite{Shishkin3}, i.e.,
 $\tilde \Sigma_{mn} \equiv \mathrm{Herm} [\langle \phi_m | \Sigma^{(i)} | \phi_n \rangle]$
and $\tilde S_{mn} \equiv \mathrm{Herm} [\langle \phi_m | \textbf{S}^{(i)} | \phi_n \rangle]$.
Then, the corresponding Hermitian eigenvalue problem
\begin{eqnarray}
\tilde U^{\dagger} \tilde S^{-1/2} \tilde H \tilde S^{-1/2}  \tilde U = \Lambda,
\end{eqnarray}
where $\tilde U$ is a unitary matrix and $\Lambda$ is the diagonal eigenvalue
matrix, in terms of which the updated wavefunctions are given as
\begin{eqnarray}
| \psi^{(i)}_n \rangle = \sum_m \tilde U_{nm} | \phi_m \rangle.
\end{eqnarray}
These updated energy eigenvalues and the wavefunctions determine the 
updated Green's function $G^{(i)}$, i.e.,
\begin{equation}
G^{(i)}(\vec x,\vec x',\epsilon)= \sum_{n}\frac{\psi^{(i)}_{n}(\vec{x})\psi^{(i)*} _{n}(\vec{x}')}{\epsilon -E^{(i)}_{n}\; \pm \mathit{i}\eta }.
\end{equation}
  The updated self-energy $\Sigma^{(i)}$ is obtained by convolving the
updated Green's function $G^{(i)}$ with the updated screened Coulomb 
interaction $W^{(i)}$, i.e., 
\begin{eqnarray}
\Sigma^{(i)}(\vec{x},\vec{x'},\omega) &=& \frac{i}{4\pi}\int_{-\infty}^{\infty}d\omega'e^{i\omega ^{'}\delta}G^{(i)}(\vec{x},\vec{x'},\omega + \omega') \nonumber \\
&\times& W^{^{(i)}}(\vec{x},\vec{x'},\omega').
\end{eqnarray} 
One finds the updated screened Coulomb interaction $W^{(i)}$ using the 
dielectric function within the random phase approximation (RPA):
\begin{equation}
W^{(i)}(\vec{x},\vec{x^{'}};\epsilon )=\int d\vec{x}^{''}\epsilon_{i}^{-1}(\vec{x},\vec{x}^{''};\epsilon )\textit{v}(\vec{x}^{''}-\vec{x}^{'}),
\end{equation}
where $\textit{v}(\vec{x}^{''}-\vec{x}^{'})$ is the bare Coulomb potential.
The dielectric matrix in the iteration $i$ is written as
\begin{equation}
\epsilon_i(\vec{x},\vec{x}^{'};\epsilon ) = \delta (\vec{x}-\vec{x}^{'})-\int d\vec{x}^{''}\textit{v}(\vec{x}-\vec{x}^{''})P^{(i)}(\vec{x}^{''},\vec{x}^{'};\epsilon ).
\end{equation}
where in the RPA we simply write that 
\begin{equation}
P^{(i)}(\vec x, \vec x'; \epsilon )=\sum_{n,m} (f_n-f_m) {{\psi^{(i)}_n(\vec x)
\psi^{(i)*}_m(\vec x)\psi^{(i)*}_n(\vec x')\psi^{(i)}_m(\vec x')}
\over {\epsilon - (E^{(i)}_m-E^{(i)}_n)}+ i \eta},
\end{equation}
where $f_n=F(E^{(i)}_n)$ and $f_m=F(E^{(i)}_m)$ and $F(E)$ is the T=0 Fermi-Dirac distribution.

\subsection{Computational Details}
All the computations were performed using the Vienna Advanced Simulation 
Package (VASP)\cite{Shishkin3,Fuchs,Shishkin2,Shishkin1}. 
The Perdew-Burke-Ernzerhof (PBE) exchange correlation 
functional\cite{PBE} was used for all GGA calculations. 
The GGA+U calculations were done using 
the Dudarev approach\cite{Dudarev} where the difference of U and J is incorporated in 
the calculation as an effective U. 
The 4s and 3d 
electrons of the transition metal atom and the oxygen 2s and 2p electrons 
were treated as valence electrons. The projected augmented wave (PAW) 
methodology\cite{Blochl} was used to describe the wavefunctions of the 
core electrons. 
The electronic 
wavefunctions were described by plane waves, where energy cutoff of 
315 eV (MnO),  500 eV (CoO), and 400 eV (NiO) were used for 
all GGA and GW calculations.

The Brillouin zone was sampled with a $4\times 4 \times 4$ k-point mesh\cite{MP76} 
and a maximum of 144 bands and 88 bands were used 
for the GGA and GW calculations. 
This size of k-point mesh is acceptable for this type of 
study, as the quasi-particle energy convergence does not depend
strongly on the k-point set\cite{VASP-GWBands}. 

\begin{figure}[htp]
\vskip 0.3 in
\includegraphics[width=3.275 in]{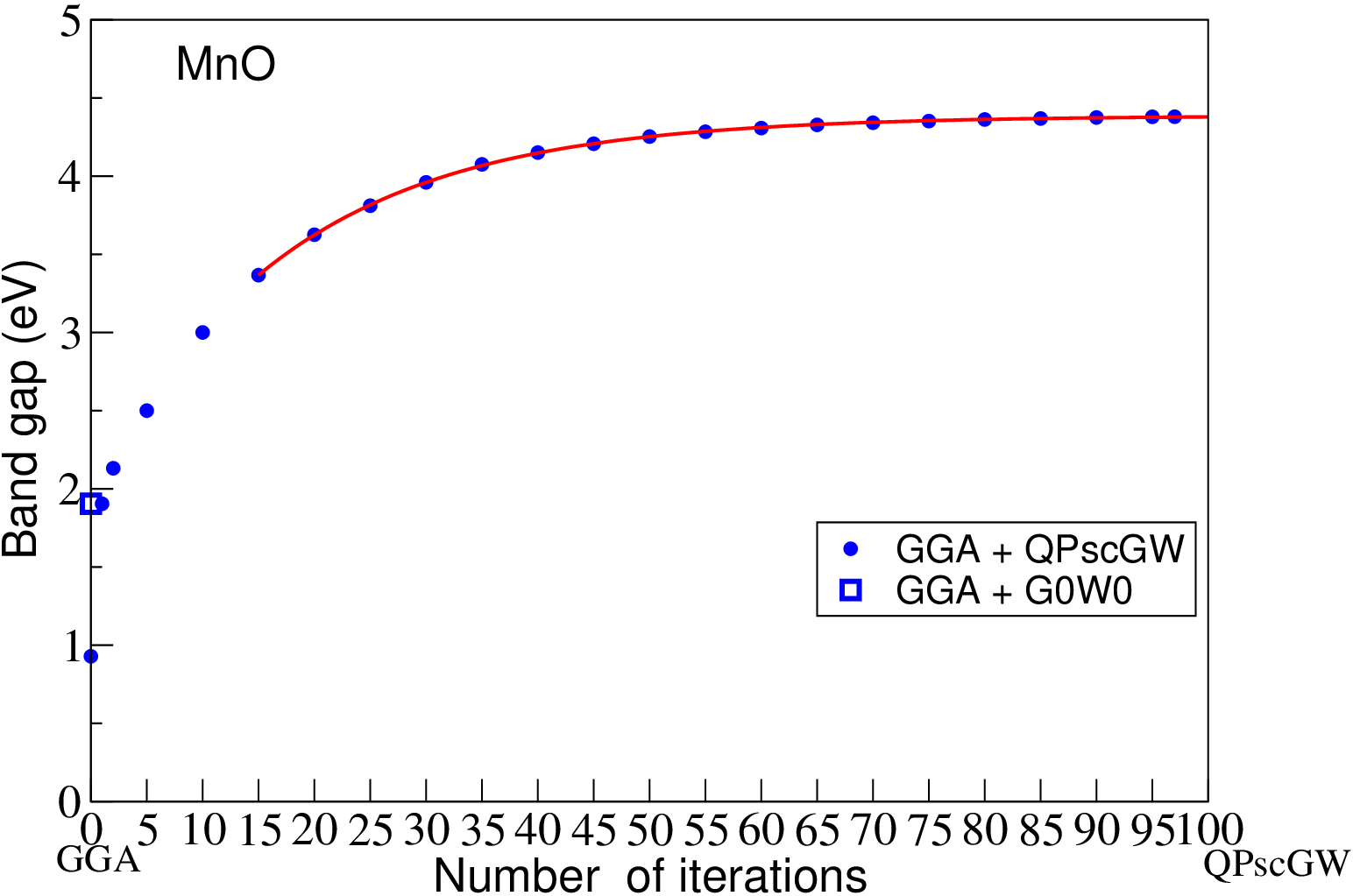}
\vskip 0.4 in
\caption{(color-online) The band-gap as a function of the QPscGW iterations.
 The starting wavefunction comes from a GGA calculation. }
\label{MnO_vs_iters}
\end{figure}

\begin{figure}[htp]
     \begin{center}
        \subfigure[]{%
            \label{fig:first-conv}
            \includegraphics[width=\figwidth]{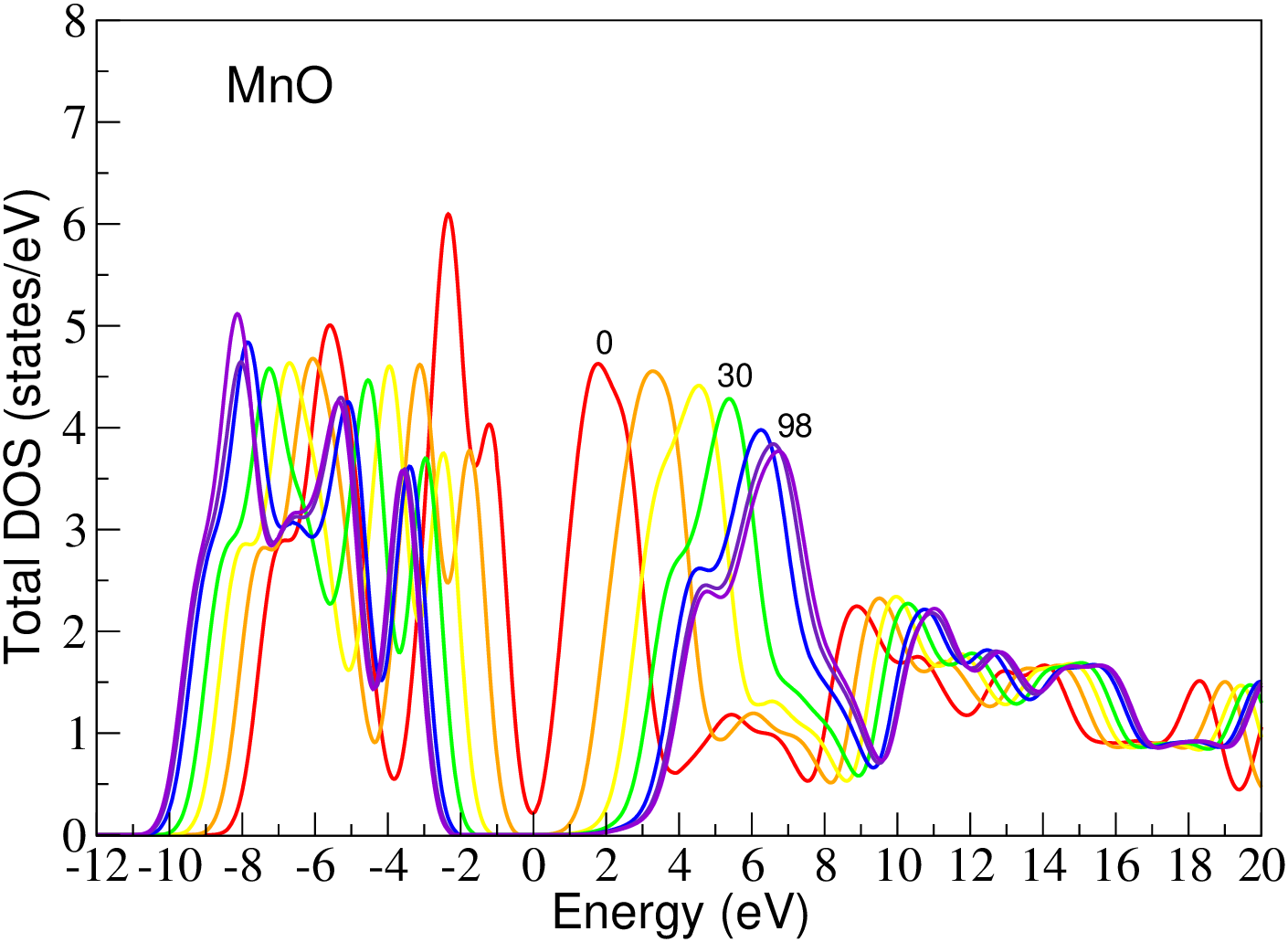}
        }
\vskip 0.3 in
        \subfigure[]{%
            \label{fig:third-conv}
            \includegraphics[width=\figwidth]{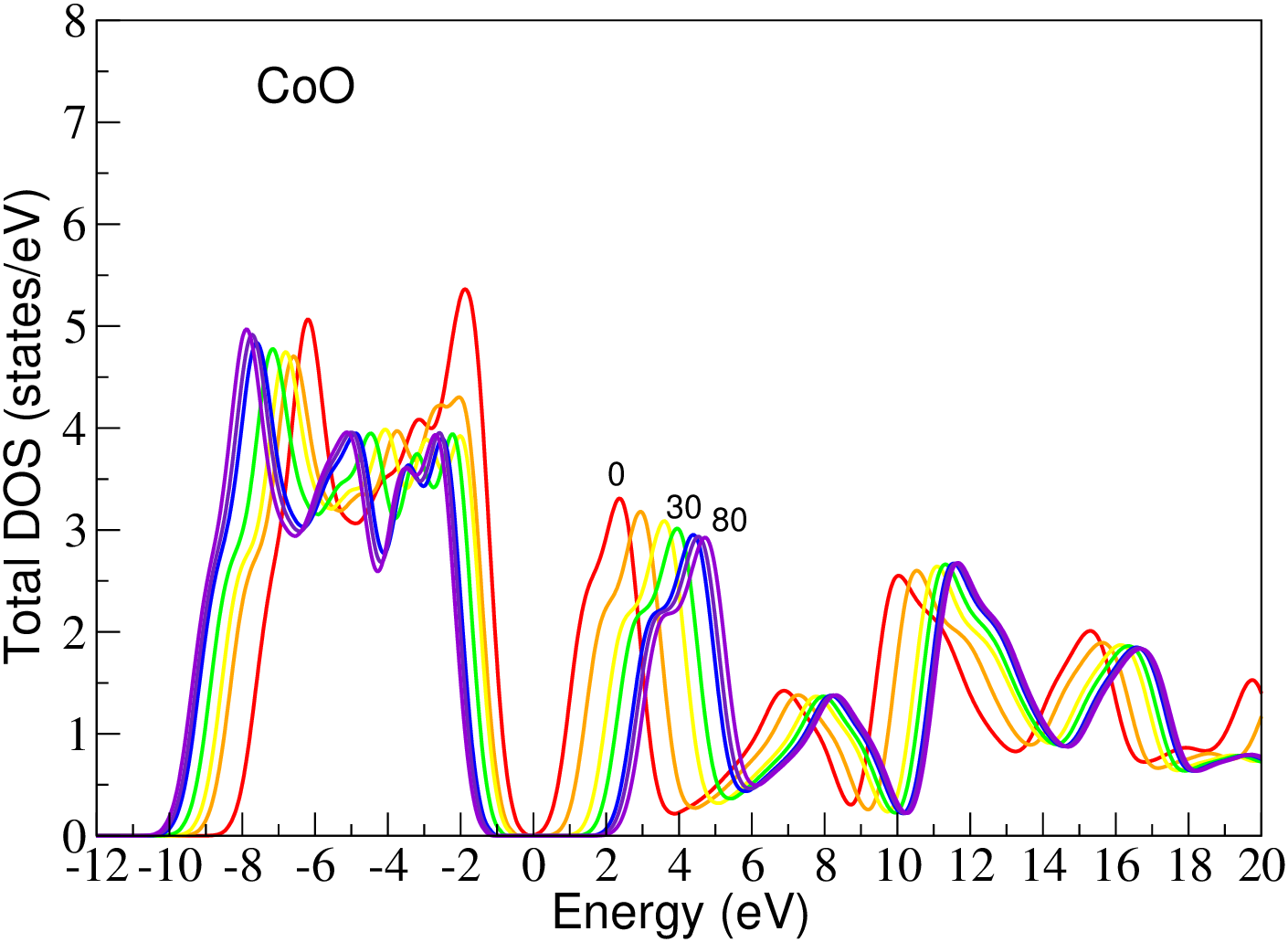}
        }
 \vskip 0.3 in        \subfigure[]{%
            \includegraphics[width=\figwidth]{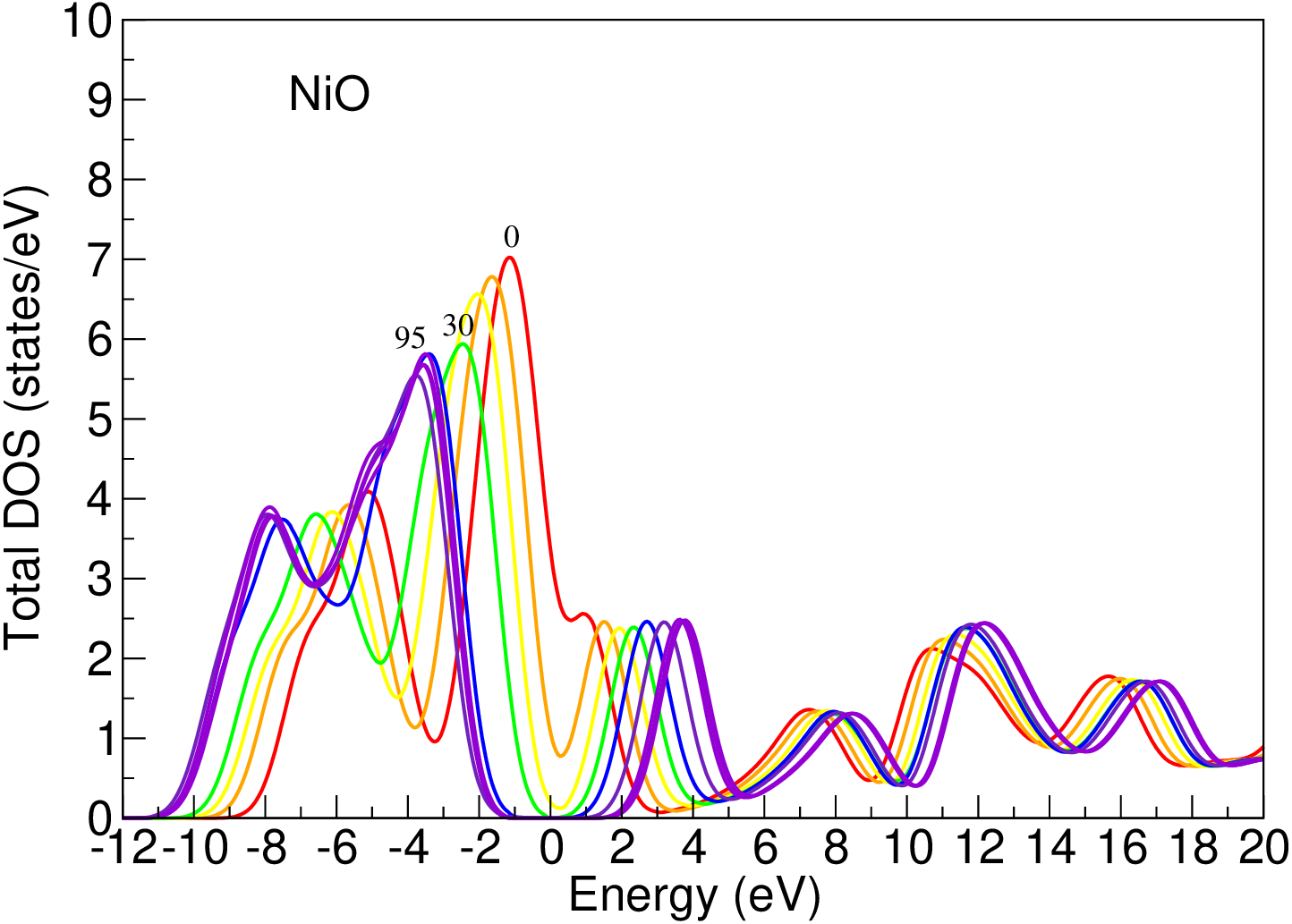}
            \label{fig:fourth-conv}
        }%
    \end{center}
\caption{(color-online) Convergence of the DOS as a function of the number of iterations
for MnO (starting from GGA), CoO (starting from GGA+U with U=3 eV),  
and NiO (starting with GGA). The curves are labeled by some selected
iteration numbers to guide the eye of the reader.}
\vskip 0.3 in
\end{figure}

Recently, the convergence of G$_0$W$_0$ calculations has
been studied with respect to number of bands and number of plane 
waves in the response function basis\cite{VASP-GWBands}.
This study also points out some convergence problems with using
non-norm-conserving pseudopotentials in the PAW methods. 
However, the current PAW pseudopotentials available are only of this type,
We can
only compare our results in this section to the results presented in
that work with non-norm-conserving pseudopotentials.
Accordingly, we have checked this using MnO and NiO as examples. 
In MnO, we find that increasing the number of conduction bands 
from 35 to 227 the direct band gap at $\Gamma$ in G$_0$W$_0$ changes 
by only 0.04~eV from 2.44~eV to 2.40~eV.
In the case of NiO, we find that
going from 66 to 478 conduction bands makes only a 0.05~eV difference
in the gap. In the aforementioned study, the difference was far more 
striking in the test material ZnO; the same amount of change
in number of bands showed a 0.2~eV difference in the gap.
We suspect ZnO to be somewhat extreme in this
regard. The previous study primarily showed change in the 
\emph{absolute} energy levels, which we are not concerned with here.
For the response-function basis set in the case of MnO by keeping 
the number of bands fixed and increasing the
energy cutoff from 200 eV ($\sim 300$ plane waves) to 
600 eV ($\sim 1500$ plane waves) we find that the direct gap at 
$\Gamma$ only changes from 2.44~eV to 2.46~eV.
All of these corrections are quite small compared to the huge changes
we find as a result of using QPscGW vs. G$_0$W$_0$.

All of the chosen materials crystallize in the rock salt structure. 
Small low temperature distortions from the cubic structure have been 
ignored in these calculations. In other work on MnO, this was 
found to have little to no effect on the electronic 
spectrum\cite{Aryasetiawan-MnO}. 
The most stable antiferromagnetic (AF II) 
state has been used, with the magnetization along alternate [111] planes. 
The lattice constants taken from experimental results as: 
8.863 $\AA$ (MnO)\cite{MnOlattice}, 
and 8.380 $\AA$ (NiO)\cite{NiOlattice}. 
For the case of CoO we used a value of 8.499 $\AA$ taken
from Ref.~\onlinecite{Bechstedt-New} which was determined such that
the volume of the (doubled) cubic unit cell to coincide with the
volume of the distorted unit cell as determined 
experimentally\cite{CoOlattice}.   
This value does not
exactly coincide with the experimental value of 
8.521 $\AA$\cite{CoOlattice} for CoO.
For the case of the other two
compounds the experimental values of the lattice constant
and those determined in order to keep the volume of the cubic unit cell
the same as the distorted unit cell are very close. 
In all three cases, we use
the rhombohedral primitive cell in our calculations. All reciprocal
lattice points are given in the basis of the corresponding
reciprocal vectors.

For the GW calculations we have used a $4\times 4 \times 4$ k-point mesh 
and a maximum of 88 bands. We used 64 
values of omega in the evaluation of the response 
functions.
\vskip 0.3 in
\section{Study of Convergence}
\label{convergence}
\subsection{ Convergence study in MnO}
\vskip 0.2 in

\begin{figure}[htp]
     \begin{center}
        \subfigure[]{%
            \label{fig:first-types}
            \includegraphics[width=\figwidth]{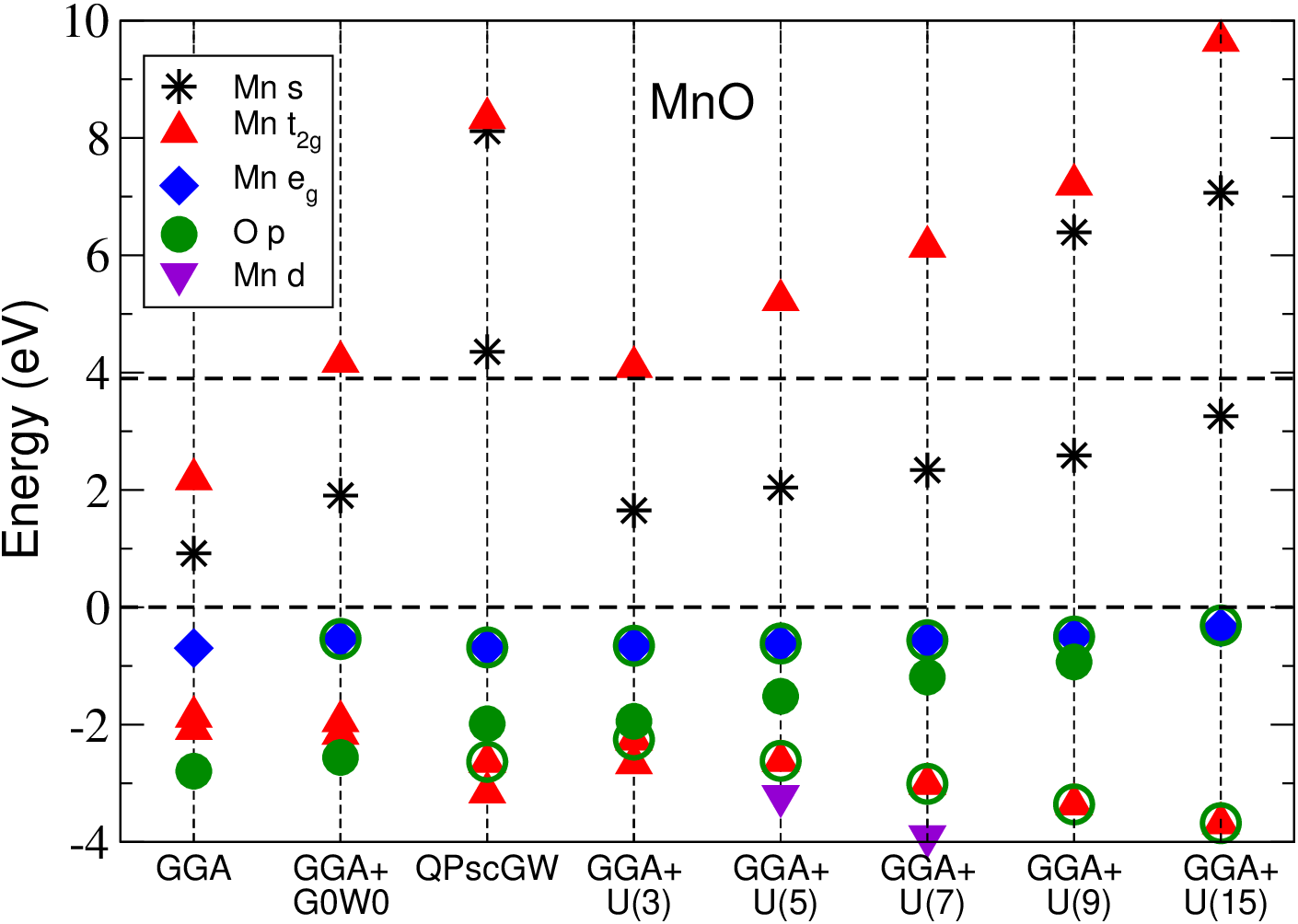}
        }
  \vskip 0.3 in   
         \subfigure[]{%
            \label{fig:third-types}
            \includegraphics[width=\figwidth]{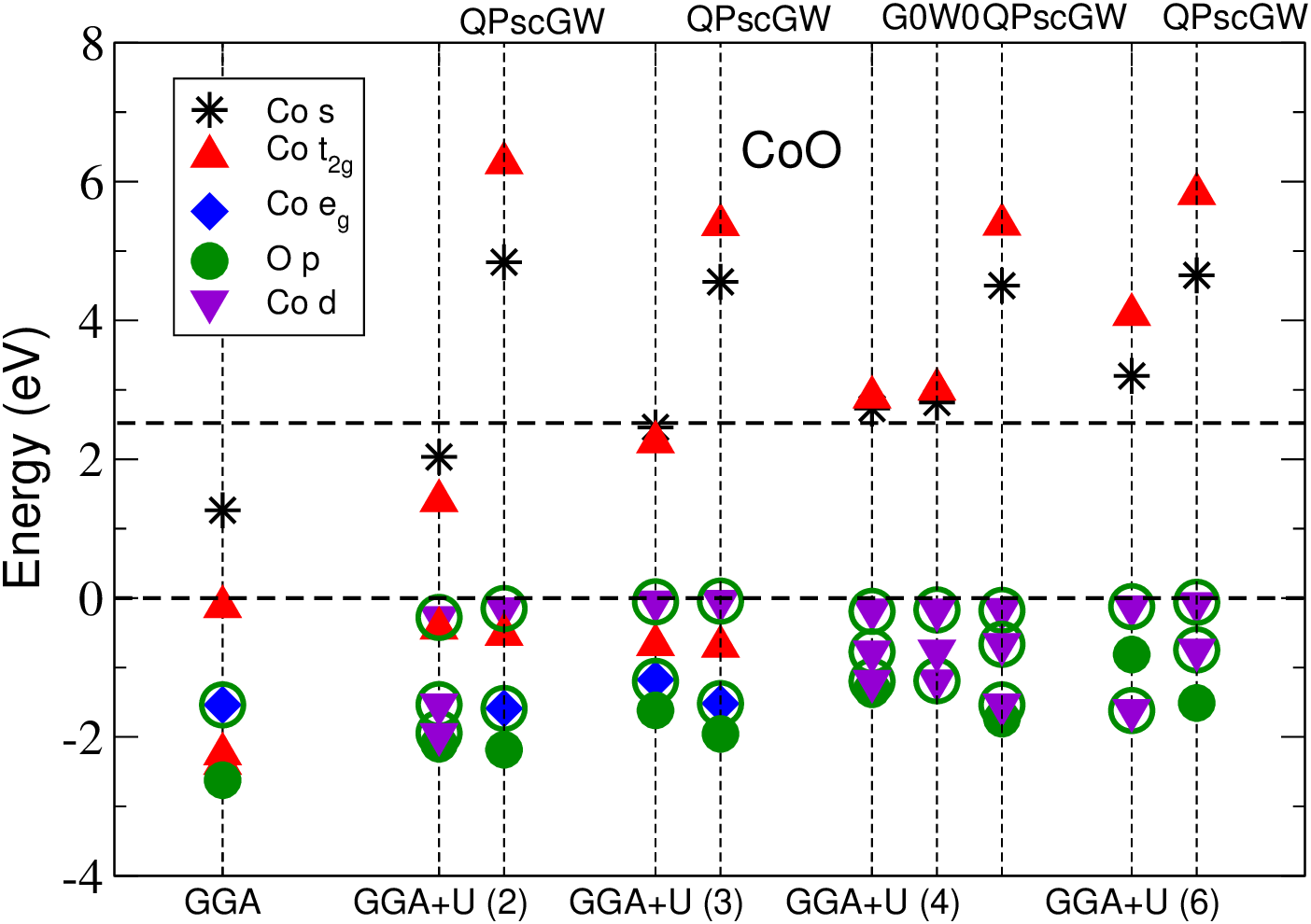}
        }
 \vskip 0.3 in        \subfigure[]{%
            \label{fig:fourth-types}
            \includegraphics[width=\figwidth]{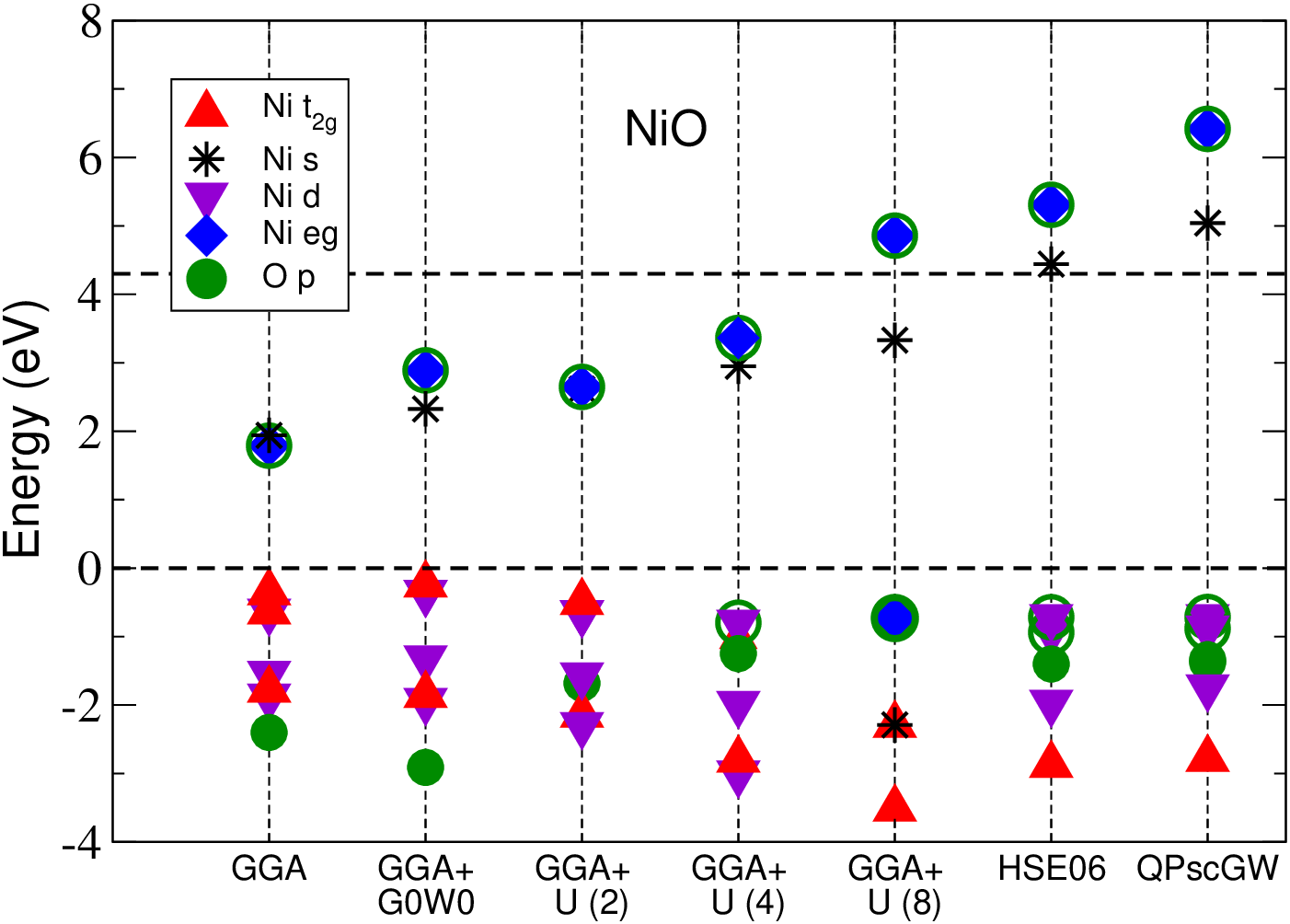}
        }%
    \end{center}
    \caption{(color-online) We present the character of the bands nearest 
   to the Fermi level at the $\Gamma$ point for 
   MnO, CoO and NiO, for different types of calculation. The various symbols 
shown in the figure legends correspond to the indicated orbital content. When multiple
symbols are used for the same band, it means that the orbital content of that particular
band is a mixture of the orbital with the corresponding overlapping 
symbols. 
Where we find clearly separated $t_{2g}$ and $e_g$ states, we mark them
accordingly; where the states of those characters are mixed, we label
them $d$ states.
We present the quasiparticle energies for a few bands above and below the ``Fermi energy''
(the valence band edge is shifted to be 0 eV) as obtained by GGA+U 
approximation with various values of U, and
by the fully iterated QPscGW calculations. The horizontal bottom and top axes are used to indicate
the approximation level. The vertical axis is used for the quasiparticle energies.}
   \label{band_types}
\vskip 0.3 in
\end{figure}

We start our QPscGW calculation with the wavefunctions obtained from an 
sGGA calculation for MnO. The convergence
of the gap as a function of iterations is shown in Fig.~\ref{MnO_vs_iters}.
Notice that the convergence is monotonic but slow, and it takes about 
80 iterations for convergence. 
We used the following simple expression
\begin{eqnarray}
\Delta_n = \Delta_{\infty} - A \exp(-{n \over {\tau}}),
\label{gap-fits}
\end{eqnarray}
to fit the dependence of each gap value $\Delta_n$ on the 
iteration $n$. The lines through the calculated points are the results
of applying this fitting procedure. This procedure gives an estimate of 
the gap extrapolated to infinite $n$, i.e., the fitting parameter $\Delta$.
For the case of MnO we find that $\Delta_{\infty}$=4.39 eV which is 
very close to the value of our last iteration (i.e., $n=98$); at $n=98$ we 
found $\Delta_{98}=4.38$ eV, which indicates that our QPscGW scheme
has converged.

We would like to stress that a single-shot G$_0$W$_0$ calculation,
shown in Fig.~\ref{MnO_vs_iters} by an open circle, is not nearly
sufficient to bridge the difference between the gap obtained within 
GGA (0.9 eV) and the converged QPscGW gap (4.38 eV). 
The G$_0$W$_0$ calculation yields a gap of approximately 
1.9 eV which is small compared to the fully converged
value of 4.38 eV. 
As we will discuss later in the case of NiO, even a G$_0$W$_0$ on top  of
HSE gives a fraction of the correction needed to bridge 
the \emph{difference} between the gap obtained at the HSE level 
and that obtained at the fully converged stage. 

Fig.~\ref{fig:first-conv} illustrates
the convergence of the density of states as a function of the QPscGW 
iteration for MnO using quasiparticle wavefunctions obtained from a 
GGA calculation. 

In Fig.~\ref{fig:first-types} we present the character of the bands
nearest to  the Fermi level for MnO and its comparison to the GGA+U 
calculations. It  will be discussed in Sec.~\ref{results} in detail.

\vskip 0.2 in
\begin{figure}[htp]
\vskip 0.3 in
\includegraphics[width=3.275 in]{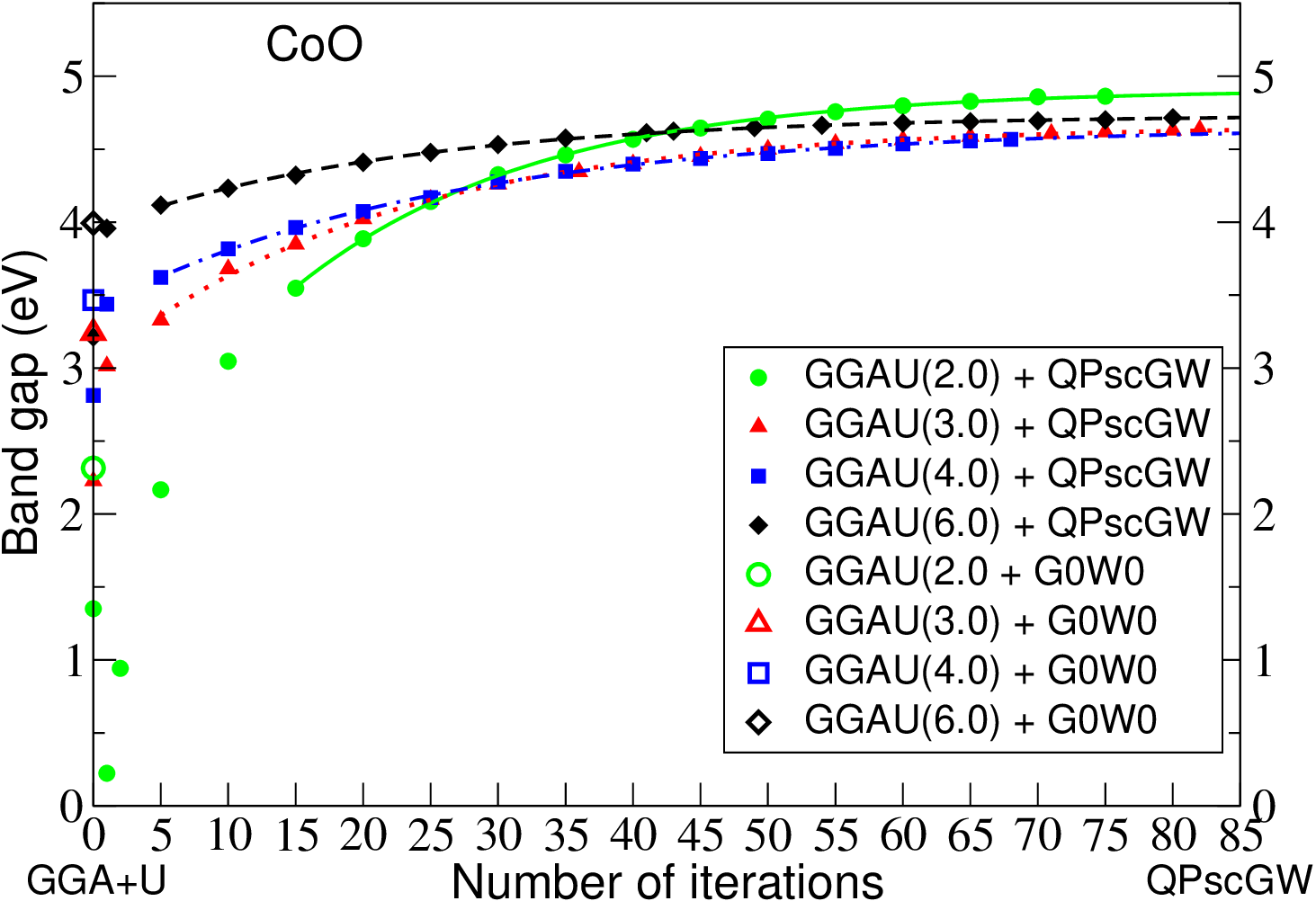}
\vskip 0.4 in
\caption{(color-online) CoO: Convergence of the band-gap as a 
function of QPscGW iterations.}
\label{CoO_vs_iters}
\vskip 0.3 in
\end{figure}

\begin{figure*}[htp]
     \begin{center}
        \subfigure[]{%
            \label{fig:CoO_bands_2}
            \includegraphics[width=\figwidth]{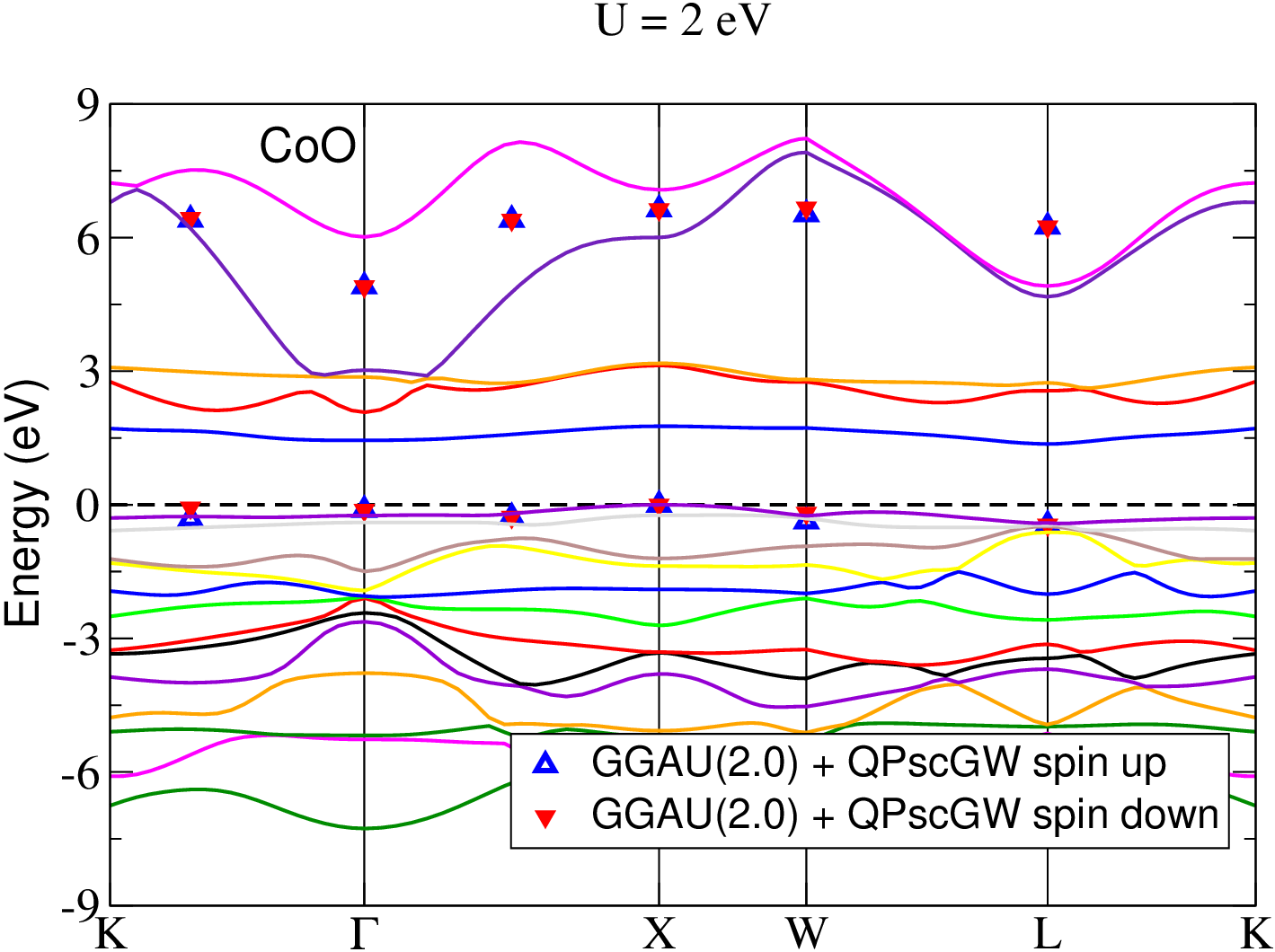}
        }%
\hskip 0.1 in        \subfigure[]{%
            \includegraphics[width=\figwidth]{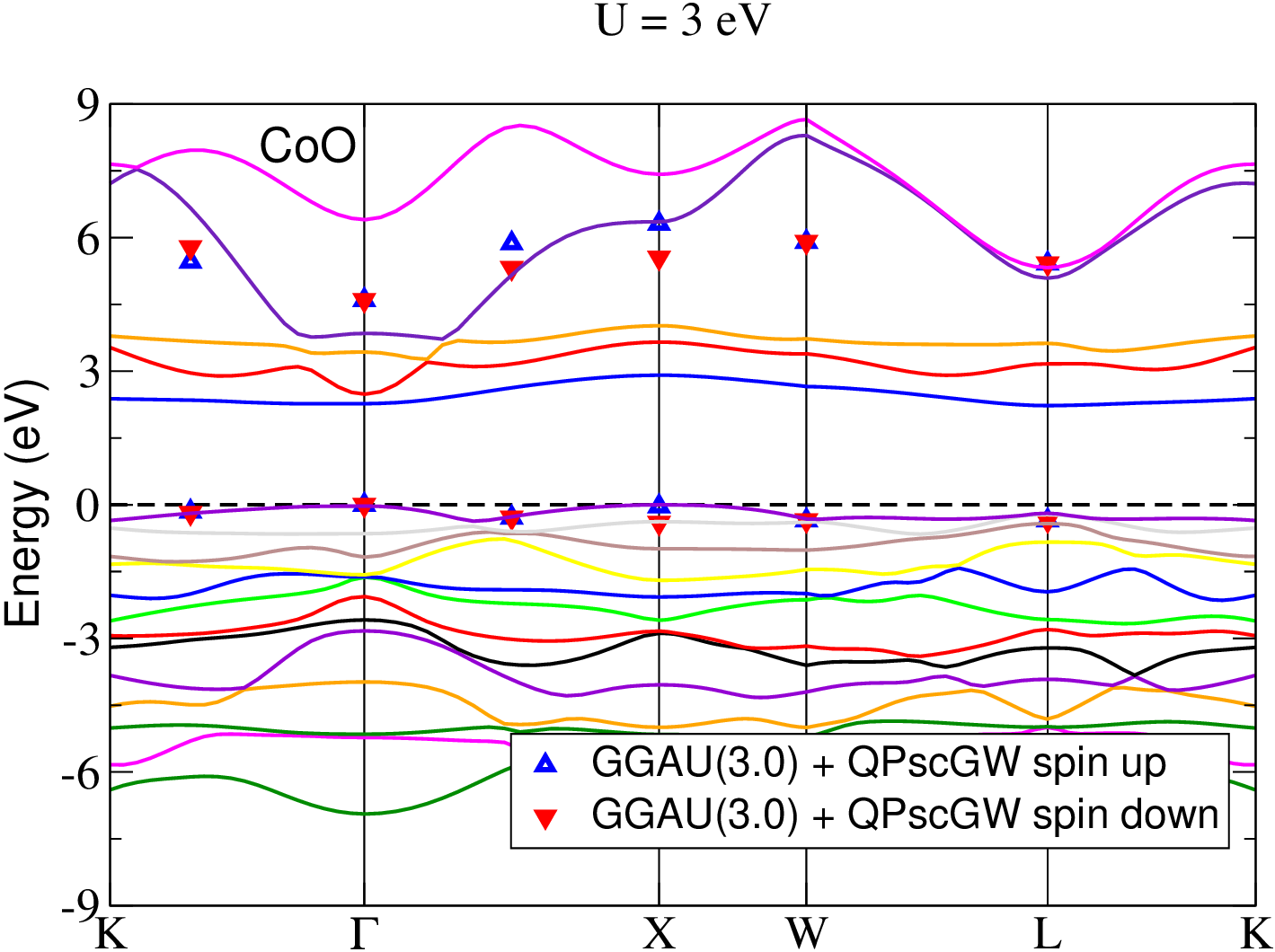}
            \label{fig:CoO_bands_3}
        }   \\
\vskip 0.3 in
        \subfigure[]{%
            \label{fig:CoO_bands_4}
            \includegraphics[width=\figwidth]{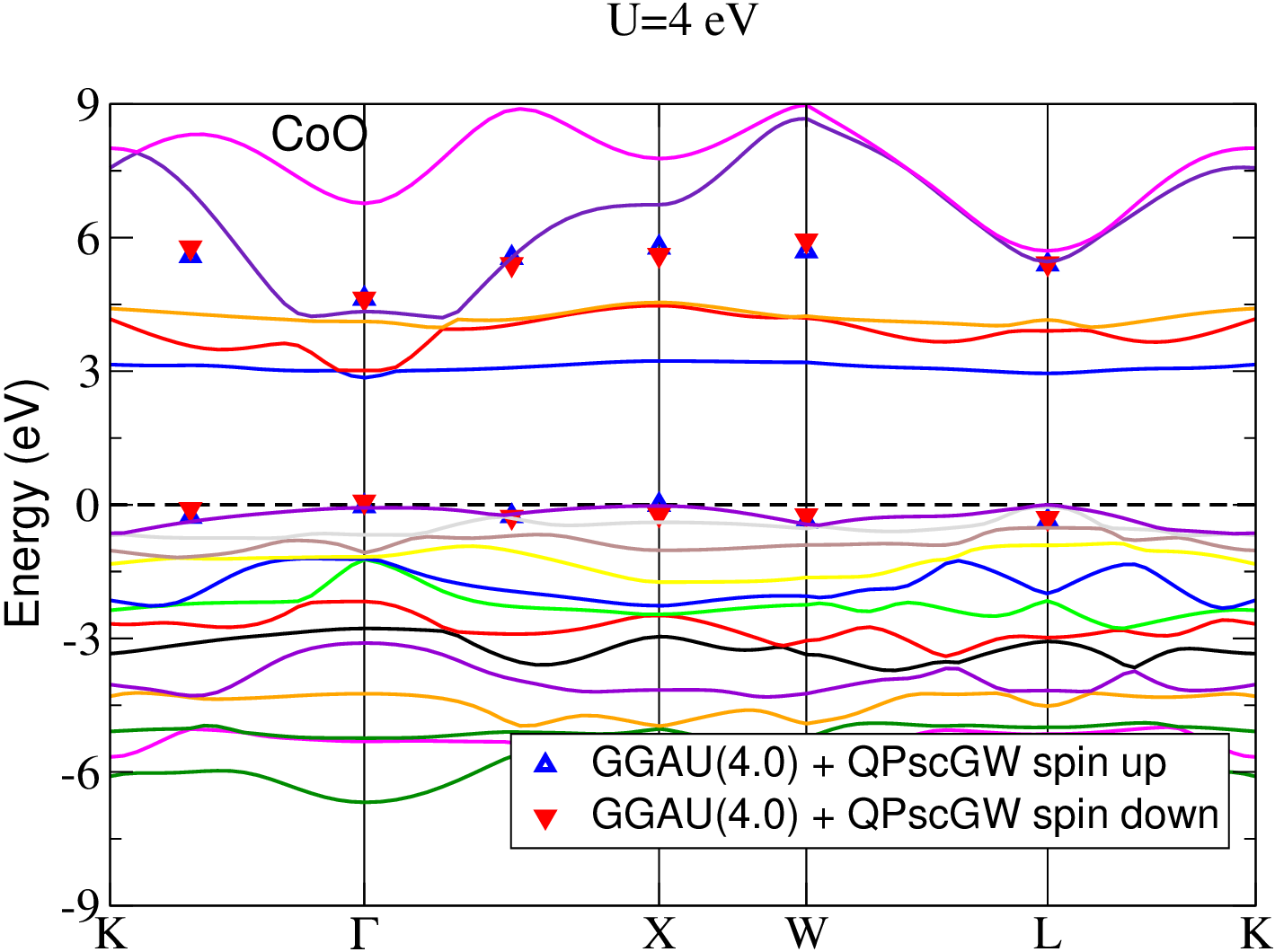}
        }%
 \hskip 0.1 in        \subfigure[]{%
            \includegraphics[width=\figwidth]{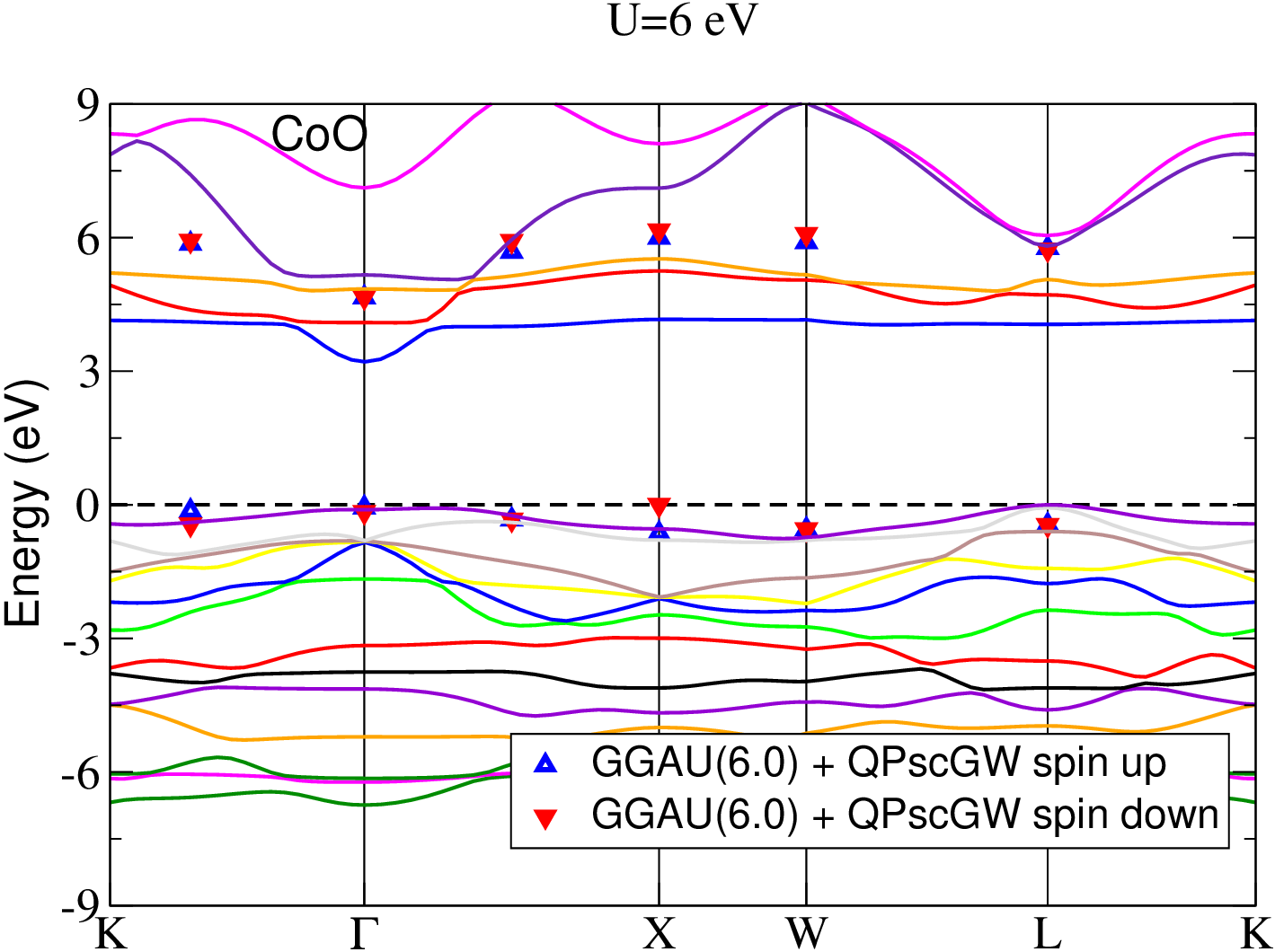}
            \label{fig:CoO_bands_6}
        }%
    \end{center}
\caption{(color-online) The energy bands of CoO near the Fermi level as obtained
from a GGA+U calculation with U=2,3,4 
and 6 eV (solid lines).
The lowest conduction and highest valence bands obtained 
from the fully converged QPscGW calculation based on the various values
of U are also shown by different symbols. 
Bands are colored only
to aid the eye; the color scheme for each material is consistent throughout.
}
\label{all-CoO-bands}
\vskip 0.3 in
\end{figure*}

\subsection{ Convergence study in CoO} 

For CoO the sGGA approach fails to yield a non-zero band-gap,
which indicates that the measured gap in CoO may be due to strong
correlations.

Let us consider the fully converged solution  $G^*$ obtained for 
the one-particle Green's 
function corresponding to the GW equation as the fixed point of an 
iterative scheme in which we start from a given $G_0$ and from any 
$G_n$ we obtain the $G_{n+1}$, etc.
This fixed point, if it exists, should be insensitive to the 
starting $G_0$, assuming that every $G_0$ used
is analytically connected to the same phase.
In order to apply many-body perturbation theory as well as reduce the 
computational cost of the GW calculation, it is important to start with 
a wavefunction close to the converged solution. 
Following these two principles in the case of CoO,
we begin the iterative QPscGW calculation
using wavefunctions obtained from a GGA + U calculation. 
We will show that because our calculation is fully converged, 
the results are only weakly dependent on the initial value of U. 
So, we perform the QPscGW calculations based on wavefunctions 
calculated with a range of values for U from 2.0 to 6.0 eV.

\begin{figure}[htp]
\vskip 0.3 in
\begin{center}
\vskip 0.4 in
\includegraphics[width=3.275 in]{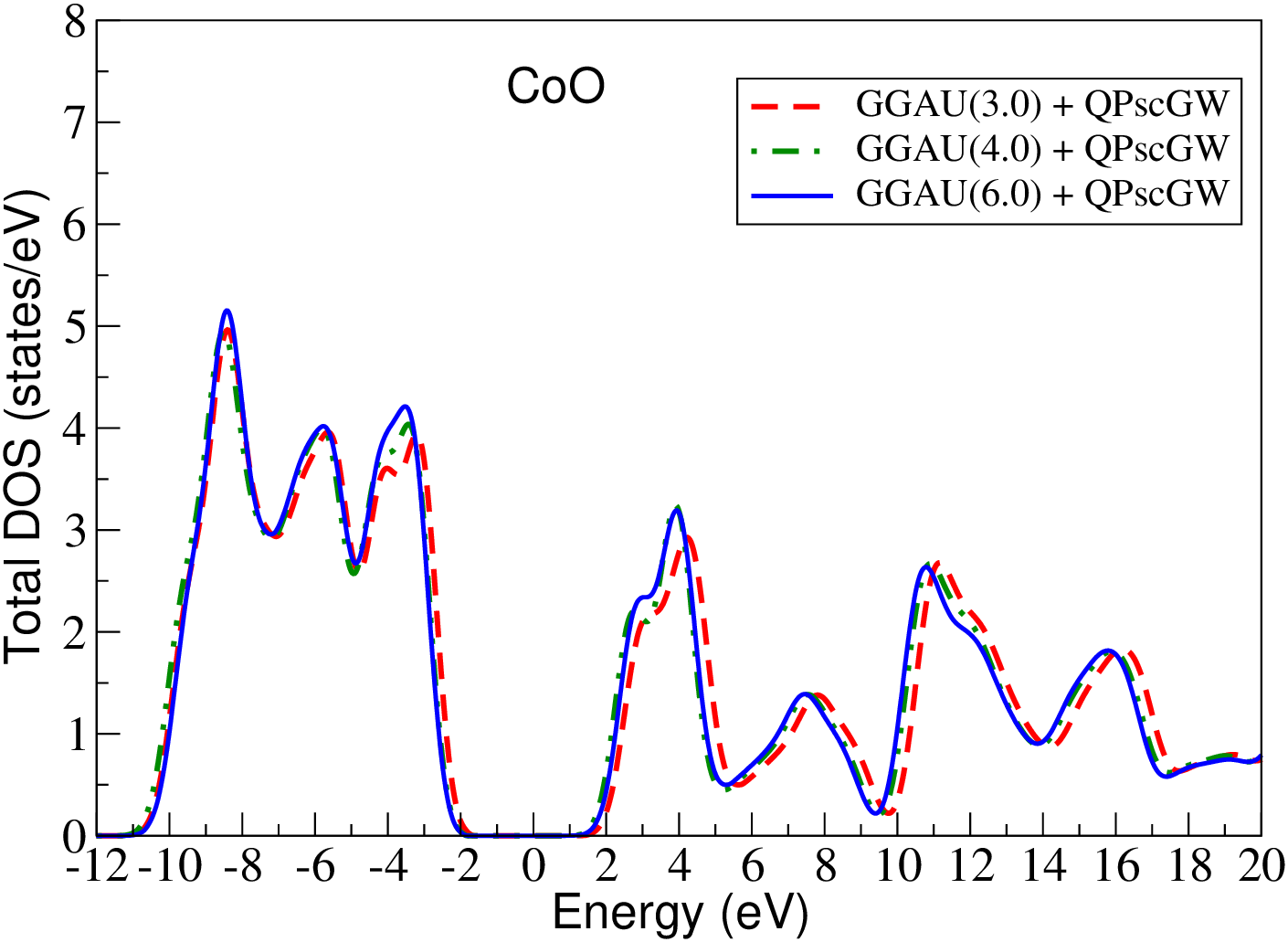}
\end{center}
\caption{(color-online)  Comparison of the DOS for the converged solutions starting from 
the results of GGA+U calculations with U=3, 4 and 6 eV.}
\label{CoO_dos_conv}
\vskip 0.3 in
\end{figure}

Recent constrained RPA calculations\cite{Sakuma,Miyake,Blugel} as well 
as other approaches\cite{Cococcioni,Pickett} suggest that the value 
of U for a simple GGA+U calculation should be around 3 eV.
The energy gap
obtained from the QPscGW procedure as a function of the iteration number is
presented in Fig.~\ref{CoO_vs_iters}. The solid lines
through the data points are fits obtained using the formula given by
Eq.~\ref{gap-fits}, which yields the following values for the extrapolated
gaps: 4.91, 4.66, 4.65, and 4.73 eV for U=2,3,4, and 6 eV respectively.
Notice that while the starting gaps for
the different values of U vary by about 2 eV, the converged values of the
gaps are about the same and fall in the range of 4.65 eV to 4.9 eV
Thus, the estimated gap is $4.78 \pm 0.13$ eV, which provides 
 an estimate of the systematic error of our QPscGW approximation in this 
case. As we will discuss below, the accumulated error from omitting vertex 
corrections, from finite size k-point mesh, and from limiting 
the number of bands, etc, we believe, is  larger than this value.

In Fig.~\ref{all-CoO-bands} we present the calculated band structure
for the four different cases. Fig.~\ref{fig:CoO_bands_2},
Fig.~\ref{fig:CoO_bands_3}, Fig.~\ref{fig:CoO_bands_4}, and 
Fig.~\ref{fig:CoO_bands_6} correspond to the results of the fully 
converged QPscGW calculation  starting from GGA+U calculations
with U=2 eV, 3 eV, 4 eV, and 6 eV respectively. 
The results of the corresponding starting GGA+U 
calculations are shown in each figure with the solid lines and
the fully converged QPscGW results for the highest occupied and the
lowest unoccupied bands are shown by up-triangles (spin up)  and 
down-triangles (spin down). 

First, notice that while the starting bands are 
very different for the four different starting GGA+U calculations,
the final QPscGW results are very close to each other.
Second,  notice that the valence band
of the QPscGW calculation is similar to the simple GGA+U calculations.
Notice also in Fig.~\ref{fig:third-types} that the ordering of the
valence bands near the Fermi energy for CoO as obtained from 
GGA+U calculations is similar to that obtained from the corresponding QPscGW 
calculations.
The third very important conclusion of this calculation is, however, 
that the conduction band obtained from the QPscGW calculation closely resembles 
the features of the GGA+U calculation for U = 4 eV and U = 6 eV. 
Namely, the results at the GGA+U level show band
crossing and hybridization in the vicinity of the $\Gamma$ point
as a function of U. 

\begin{figure*}[htp]
     \begin{center}
        \subfigure[]{%
            \label{NiO_gap_vs_iters_up}
            \includegraphics[width=\figwidth]{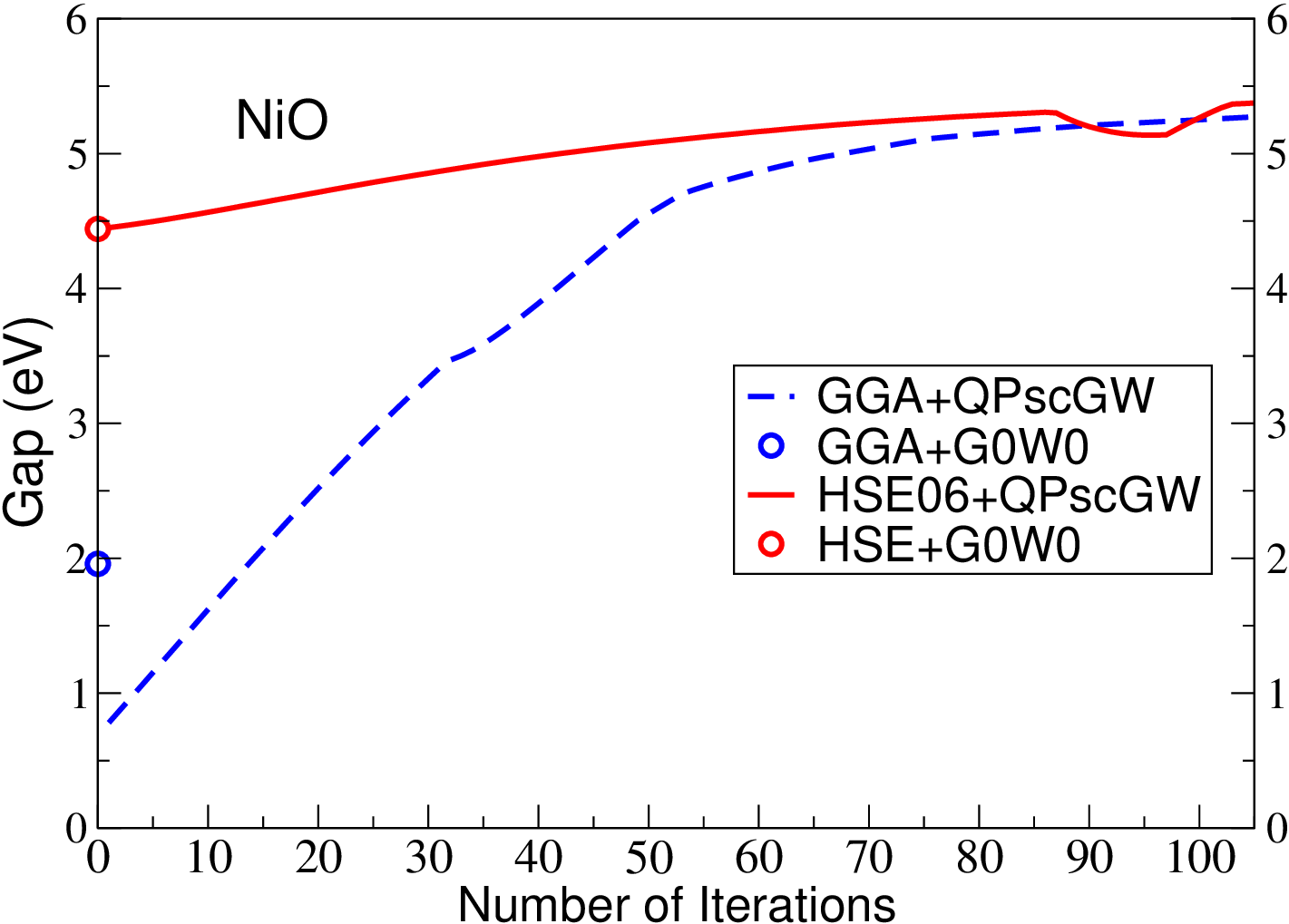}
        }%
  \hskip 0.1 in       \subfigure[]{%
            \label{NiO_gap_vs_iters_down}
            \includegraphics[width=\figwidth]{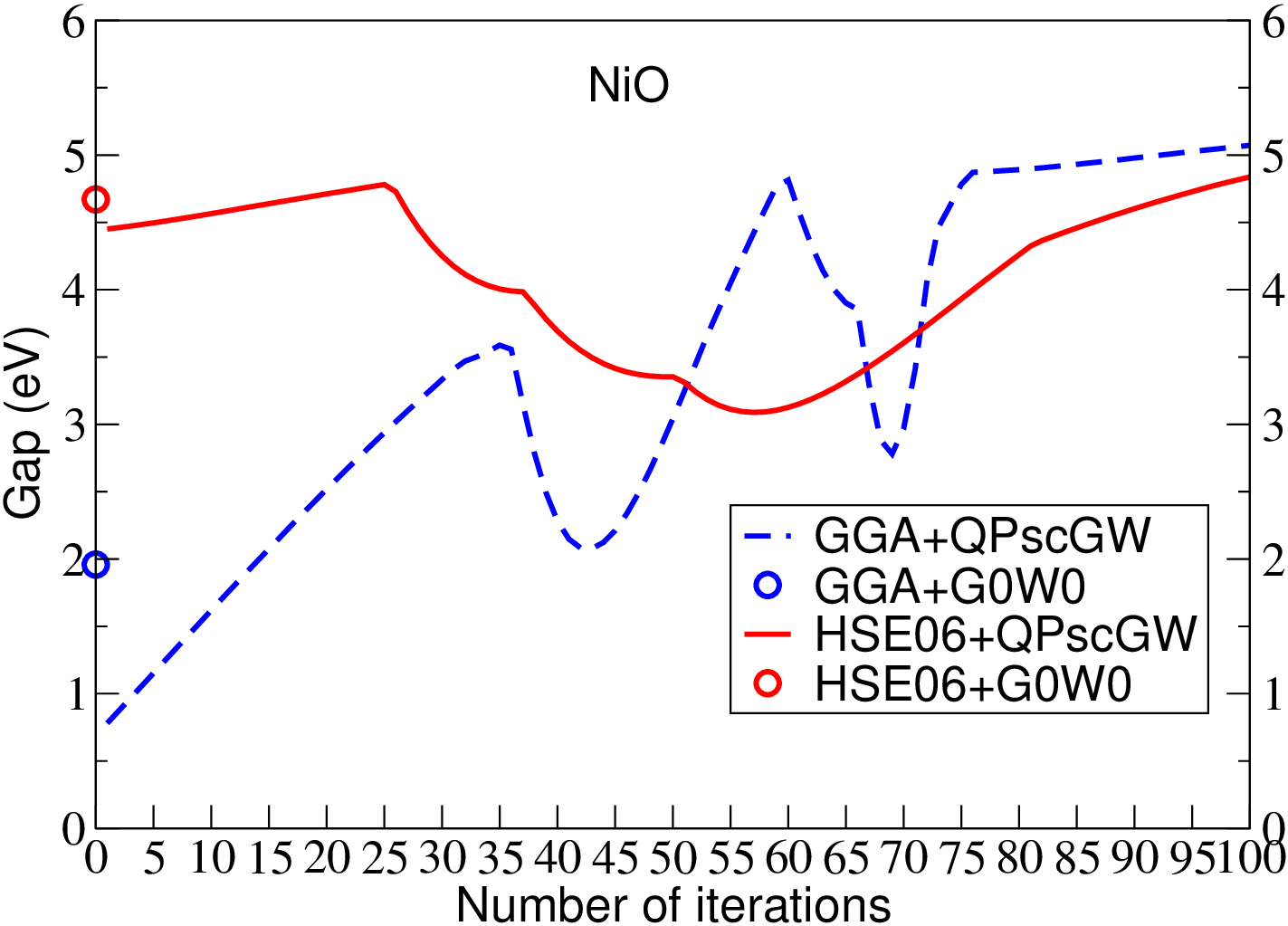}
        } \\
\vskip 0.3 in
        \subfigure[]{%
            \label{NiO_DOS_SCGW}
            \includegraphics[width=\figwidth]{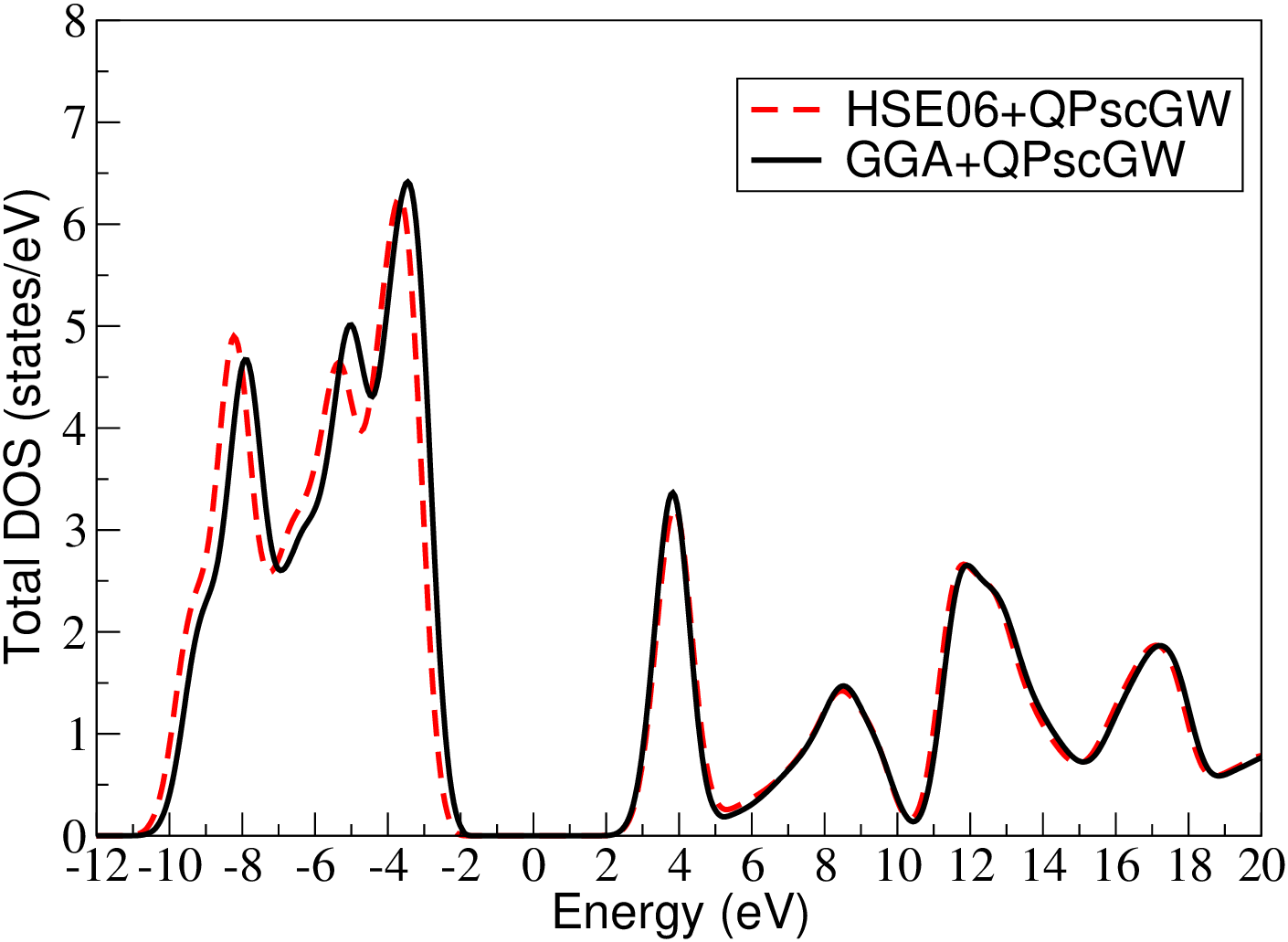}
        }%
  \hskip 0.1 in       \subfigure[]{%
            \label{NiO-level-crossing}
            \includegraphics[width=\figwidth]{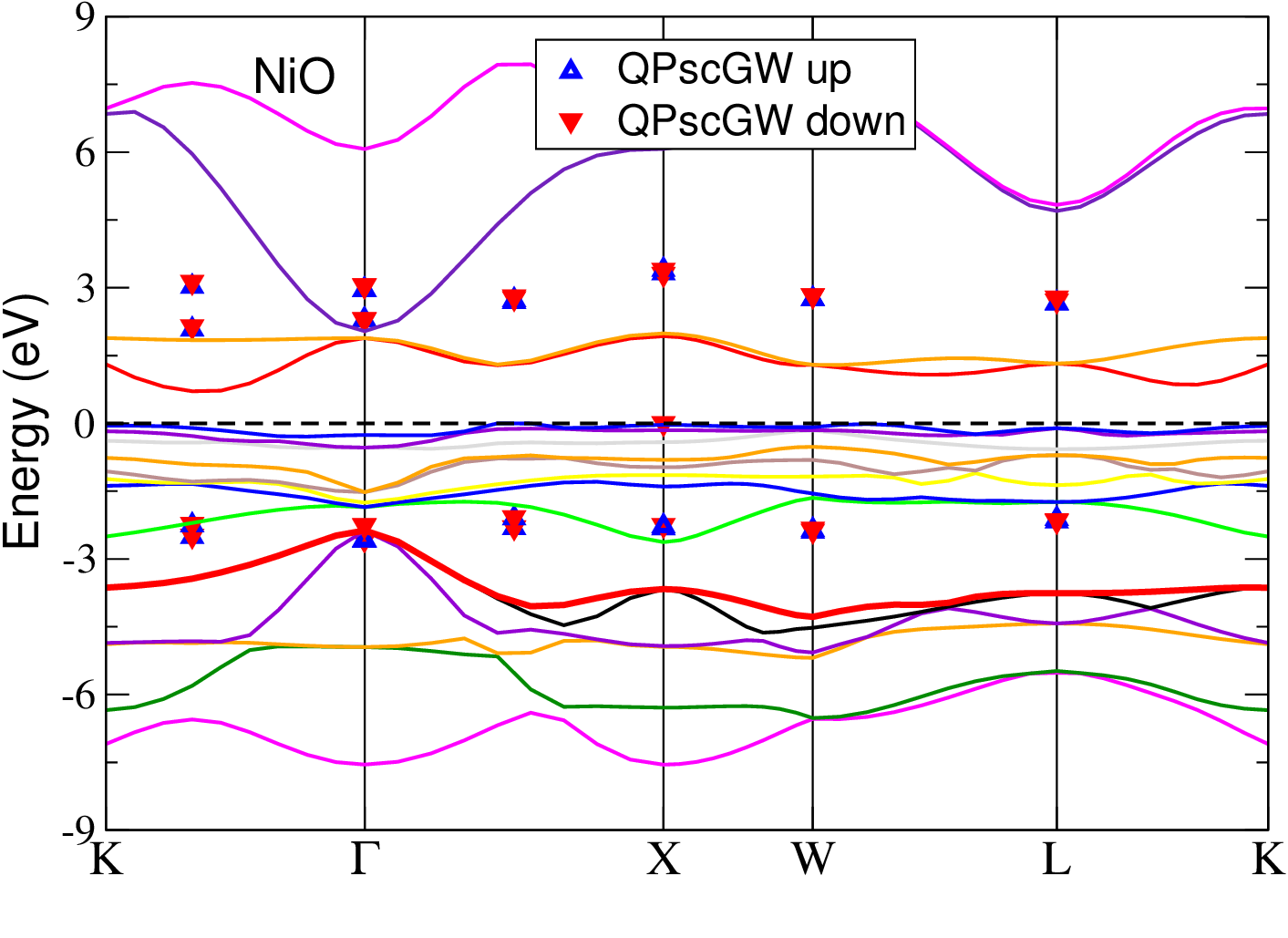}
        }%
    \end{center}
\caption{(color-online) NiO: (a) Convergence of the band gap for spin-up electrons 
with iterations starting from GGA and HSE wavefunctions.
(b) Convergence of the band gap between spin-up and spin-down electrons 
with iterations starting from GGA and HSE wavefunctions.
(c) Comparison of the converged density of states obtained after
QPscGW calculations starting from HSE06 and sGGA.
(d) Illustration of the ``level crossing" state at the 43rd step
of the QPscGW procedure in which a spin-down state temporarily moves to a
higher energy, decreasing the gap.
Bands are colored only
to aid the eye; the color scheme for each material is consistent throughout.
}
\vskip 0.3 in
\end{figure*}

\begin{figure}[htp]
     \begin{center}
        \subfigure[]{%
            \label{fig:first-bands}
            \includegraphics[width=\figwidth]{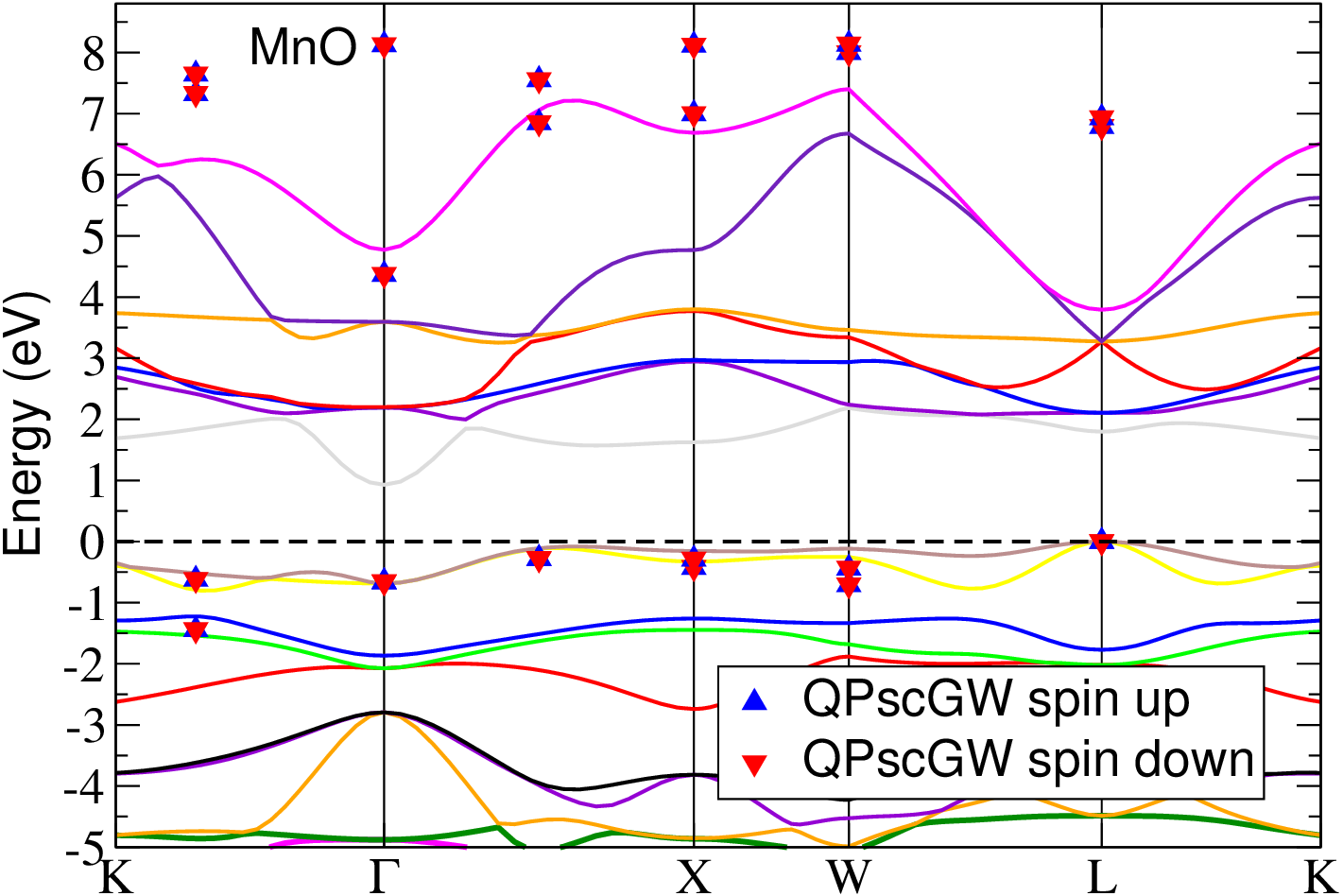}
        }
\vskip 0.35 in    
        \subfigure[]{%
            \label{fig:third-bands}
            \includegraphics[width=\figwidth]{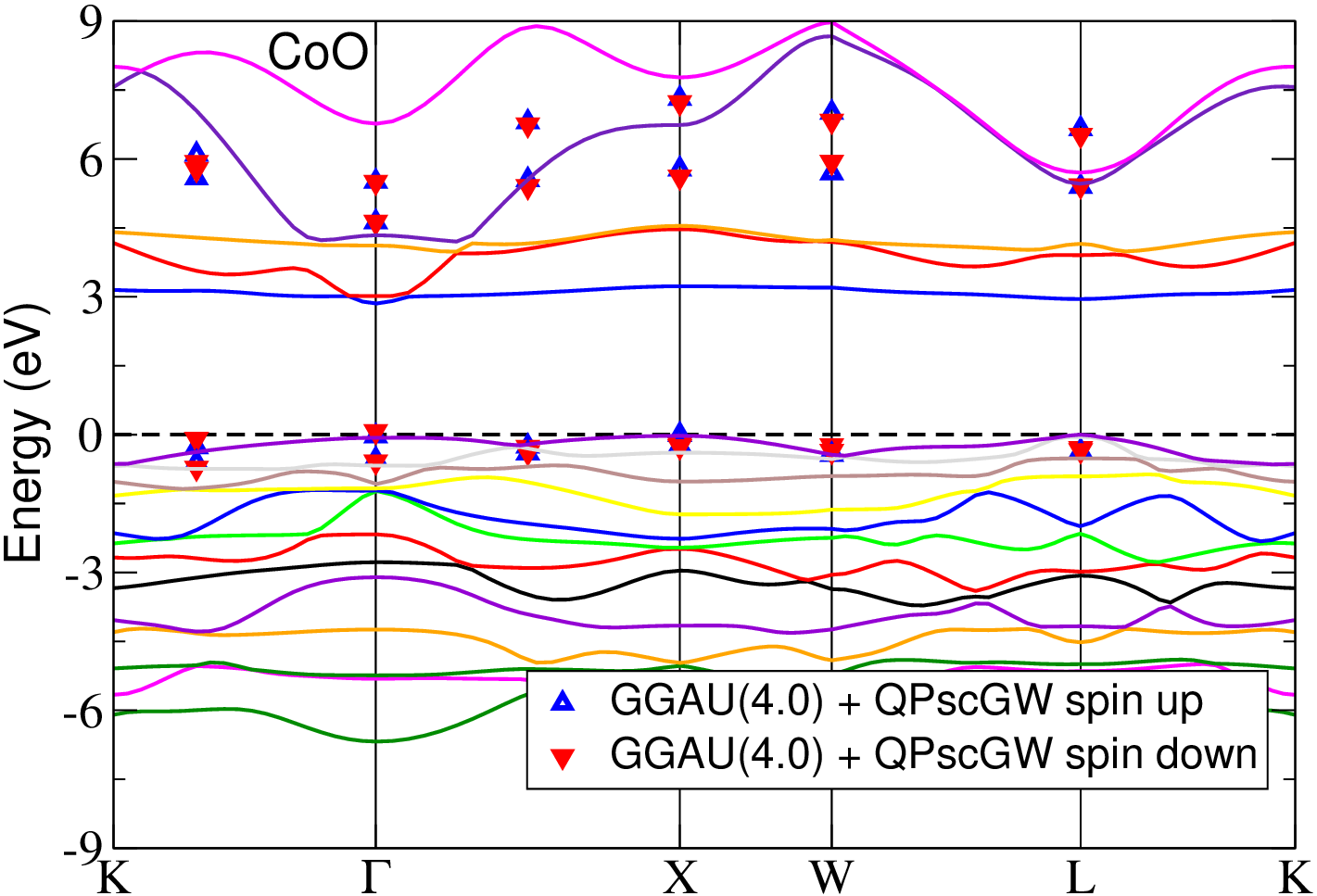}
        }
 \vskip 0.35 in        \subfigure[]{%
            \includegraphics[width=\figwidth]{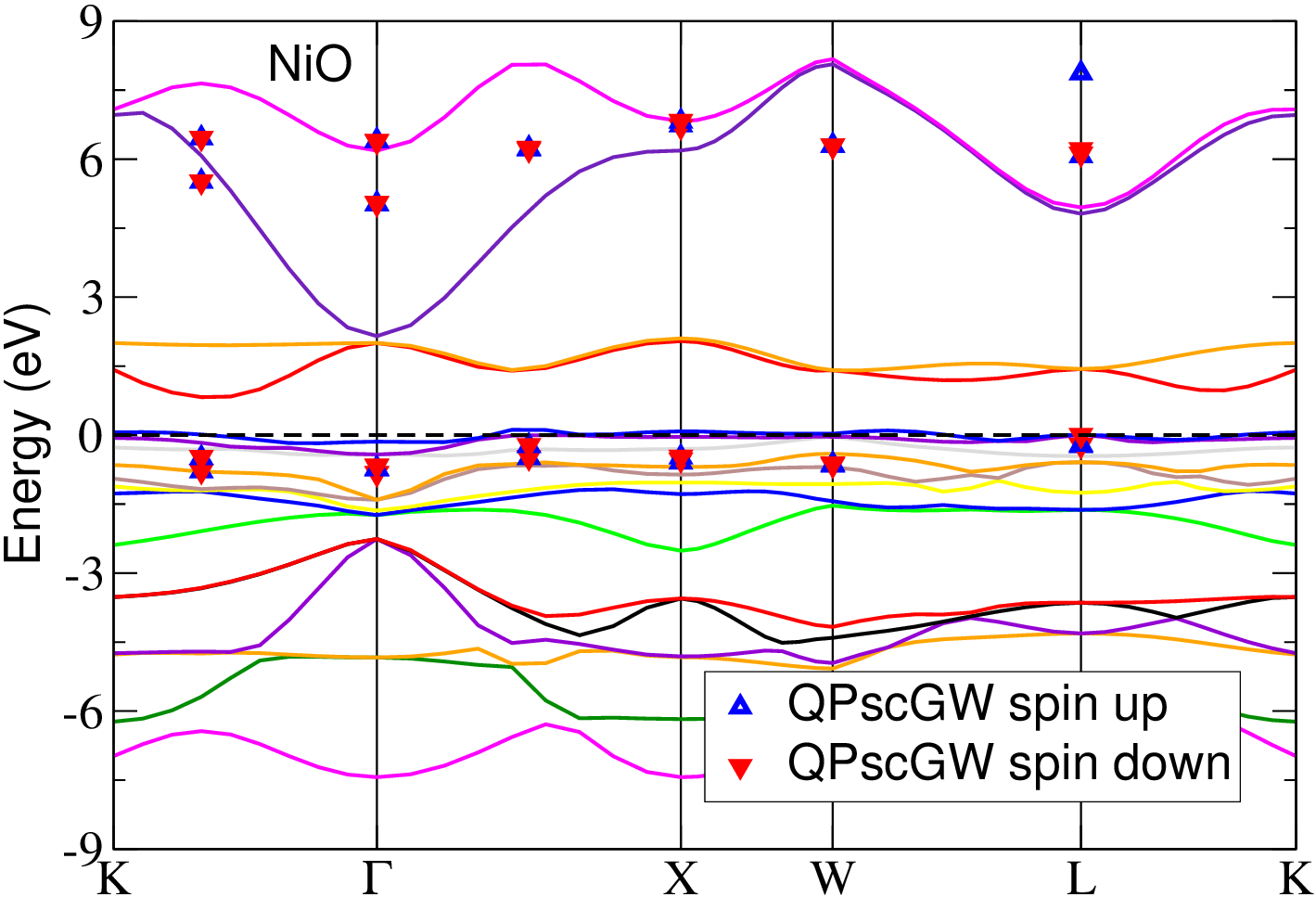}
            \label{fig:fourth-bands}
        }%
    \end{center}
\caption{(color-online) The bands obtained (a) for MnO using sGGA (solid lines)
and fully converged QPscGW (open or full circles),
(b) CoO using GGA+U (solid lines)
and fully converged QPscGW (open or full circles),
and (d) for NiO using sGGA (solid lines)
and fully converged QPscGW (open or full circles)  
Bands are colored only
to aid the eye; the color scheme for each material is consistent throughout.
}
\label{all-bands}
\vskip 0.3 in
\end{figure}
\begin{figure}[htp]
     \begin{center}
        \subfigure[]{%
            \label{fig:first}
            \includegraphics[width=\figwidth]{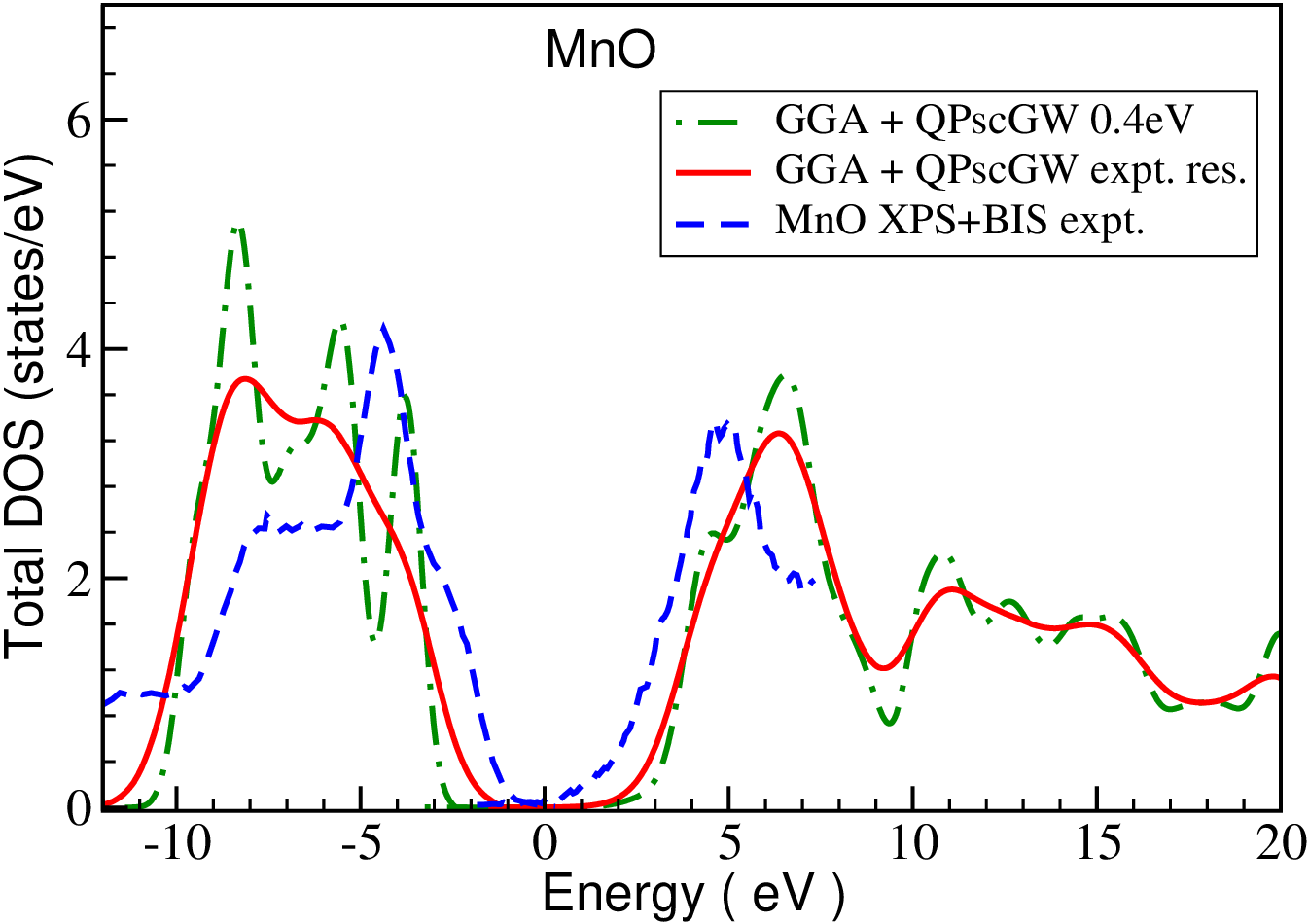}
        }
 \vskip 0.35 in  
         \subfigure[]{%
            \label{fig:third}
            \includegraphics[width=\figwidth]{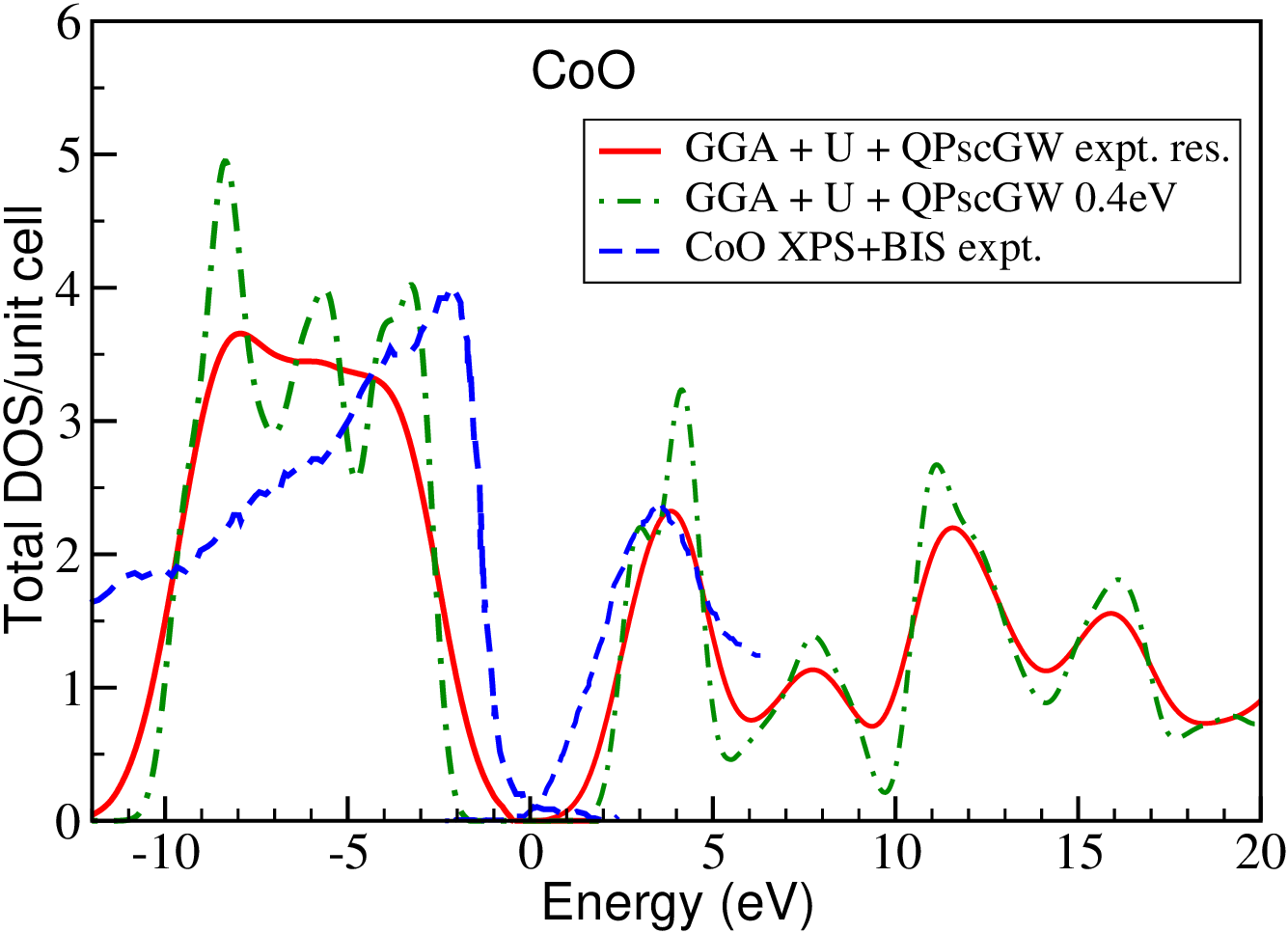}
        }
 \vskip 0.35 in        \subfigure[]{%
            \label{fig:fourth}
            \includegraphics[width=\figwidth]{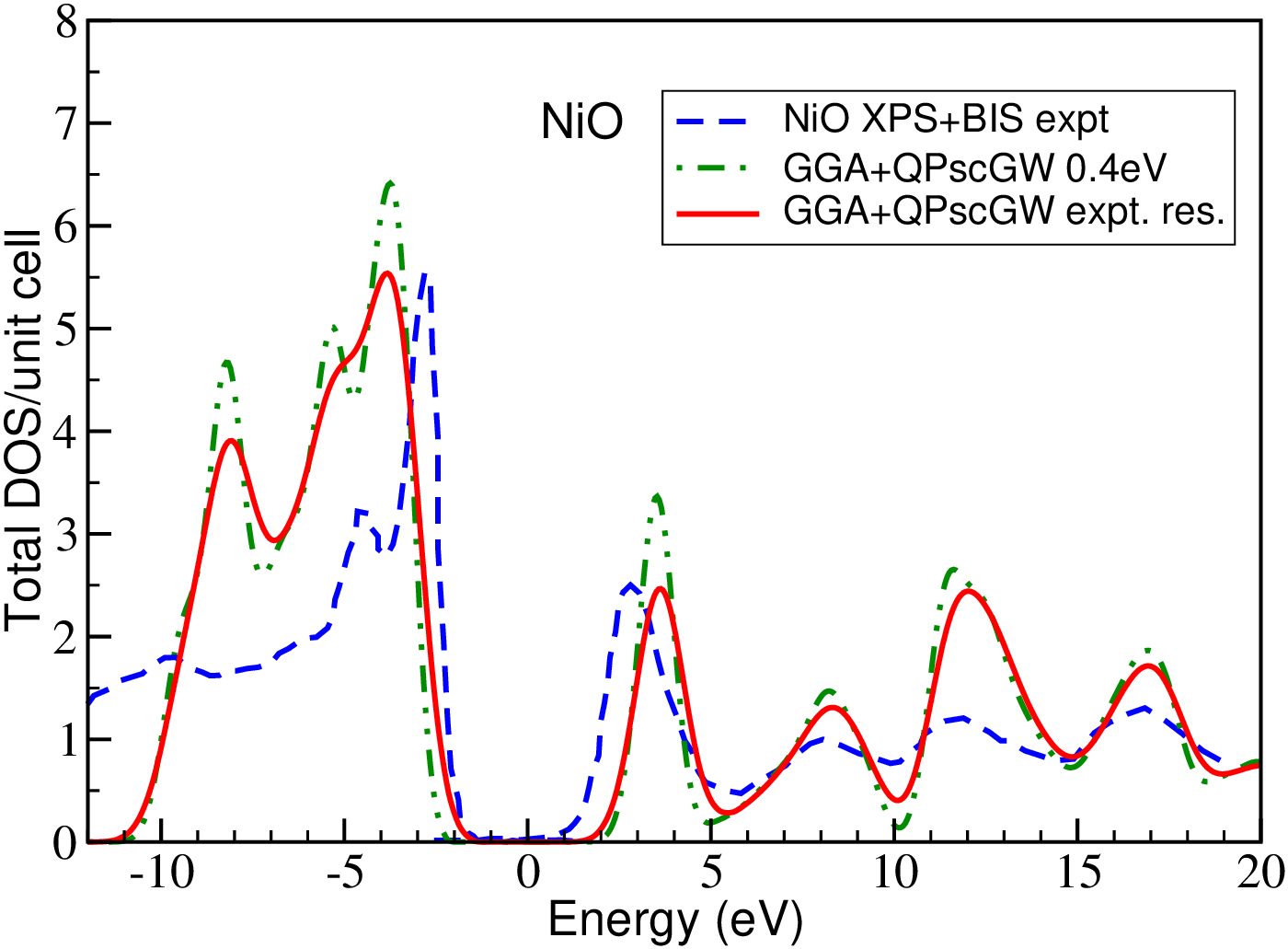}
        }%
    \end{center}
    \caption{(color-online)The density of states of transition metal oxides MnO,  
NiO,  and NiO  obtained by solving the GW equations self-consistently is 
compared to the experimental XPS and BIS data.\cite{Berger, Elp, Allen} }
   \label{dos_vs_expt}
\vskip 0.3 in
\end{figure}

The hybridization starts at U$\sim 4$ eV and involves
one dispersionless lower energy band of Co $t_{2g}$ character
and one higher energy, strongly dispersive Co $s$ band.  
The result of the hybridization is to produce
a conduction band which is dispersive only near the $\Gamma$ point where
the hybridized lower energy band inherited the dispersion from 
the $s$ band. This behavior also occurs
in the transition at around U = 4 eV seen in the
level ordering  of the lowest two conduction bands in 
Fig.~\ref{fig:third-types}.
This is the exact same character as the lowest conduction band 
obtained as the final result of the QPscGW iterations independently
of the starting wavefunctions.  
Namely, the result of the QPscGW calculations and that of
the simple GGA+U calculations above U$\sim 4$ eV are qualitatively
very similar and the main difference is in the size of the gap.

It is important to discuss some additional details of our results.
Notice that the energies of the converged QPscGW states having different
spins, with all other quantum numbers common, are somewhat different.
The reason is that the energy levels for up and down spins
are free to vary independently in the present QPscGW calculations. As we will
discuss in the case of NiO this effect becomes more pronounced there.
The departure from up-down symmetry is temporarily 
explored  by the QPscGW iteration procedure as a possible path to reaching 
a fully converged solution; however, it seems that the final fully 
converged solution is the up-down symmetric one. 

We have noticed that during the QPscGW evolution from the starting to the 
final state, level crossings have occurred in all four cases. 
Starting with U=3 eV, for example,
the conduction band minimum for spin-up, obtained from the initial 
GGA+U calculation, is at $\vec k=(1/2,1/2,1/2)$ (in units of the reciprocal 
lattice vectors of primitive unit cell of the rock-salt structure), 
while the 
converged QPscGW calculation yields a conduction band minimum at 
the $\Gamma$ point. Furthermore, in general, we find that the orbital 
ordering, as well as the exact $\vec k$ point where the valence band 
maximum occurs, may be slightly different for the 
converged states for different values of U. These discrepancies between 
the character of the converged states starting from different values of
U may be due to the following reasons:

a) The lowest conduction band, as well as the highest valence band, 
have a narrow bandwidth, 
consistent with the belief that CoO is a strongly correlated 
material with nearly localized electrons near the
Fermi level. Therefore, the energy eigenvalues 
corresponding to various values of $\vec k$ are different by a very small
amount which may be beyond the level of accuracy of the present 
QPscGW calculation.  This is illustrated in Fig.~\ref{all-CoO-bands}.

b) When level crossing occurs, during the QPscGW evolution,
 the adiabatic nature of such an evolution can not be guaranteed. Namely, 
the adiabatic theorem states that:  
in a real-time evolution is adiabatic,
i.e.,  in order for the eigenstates found by solving the time-dependent
Schr\"odinger equation to  be in one-to-one correspondence 
with the starting eigenstates and to correspond to the eigenstates
of the perturbed instantaneous Hamiltonian there must be 
no level crossing during the time-evolution.

c) We found that already, at the GGA+U level,
there is such a level crossing as a function of $U$. For example, the 
valence band maximum for U=2 eV occurs
at $\vec k=(0,1/2,1/2)$, while for  U=6 eV it occurs 
at $\vec k=(1/2,1/2,1/2)$.  Therefore, starting
from unperturbed Hamiltonians corresponding to such different values
of U, the QPscGW is forced to evolve through at least one 
level crossing in order to reach the same solution at the converged stage.
Fig.~\ref{all-CoO-bands} demonstrates the band crossing 
which happens at the GGA+U level. The lowest conduction band,
which is nearly dispersionless and of Co $t_{2g}$ character 
at $\Gamma$, hybridizes near $\Gamma$ with another band of pure $s$ 
character, as we increase the value of U from 3 to 4 eV.

More general features
such as the converged QPscGW density of states, obtained with 
different starting wavefunctions, agree with each other quite nicely.
Fig.~\ref{fig:third-conv} illustrates
the convergence of the density of states as a function of the QPscGW 
iteration for CoO using quasiparticle wavefunctions obtained from 
a GGA+U calculation with U = 3.0~eV.
Here, we begin with a gap of 2.23 eV at the GGA+U level with U = 3.0~eV. 
After 80 iterations, where convergence is achieved, and using the 
extrapolation discussed in the previous subsection we obtained a 
gap of 4.66~eV.

In Fig.~\ref{CoO_dos_conv}, we show that starting with wavefunctions 
obtained from different GGA+U calculations with 
U = 3.0, 4.0 and 6.0 eV the 
QPscGW procedure converges to very similar density of states.
Although the GGA+U band gaps with U = 3.0, 4.0 and 6 eV differ
by 2.0 eV, the final QPscGW calculations converge to similar density of
states and similar gaps with a spread of $\pm0.13$ eV. 
This indicates that these QPscGW calculations are to a certain degree 
weakly dependent on the choice of the starting wavefunctions 
within the regime of perturbation theory.

\subsection{ NiO: Starting with GGA, and HSE}

It has been claimed that although QPscGW results may be preferable, they 
are computationally very costly, thus, using starting wavefunctions 
obtained from hybrid functionals such as the HSE 
\cite{HSE}, followed by a single shot GW calculations may be a good 
practical alternative\cite{Rodl-G0W0+HSE}.  
However, the suitability of this approach for TMOs was recently 
questioned\cite{Coulter-scGW}, namely, whether or not  we can just 
stop at low order in a GW approach having started the GW calculation from 
wavefunctions and quasiparticle energies obtained from such an HSE calculation.

In Fig.~\ref{NiO_gap_vs_iters_up}, we present the results of two
different QPscGW calculations, one
starting from GGA and a second starting from the HSE06 functional.
This figure shows the gap for up-spin states. 
First, it appears that both calculations converge smoothly.
Notice that the result of the G$_0$W$_0$ on top of either GGA or
HSE06 makes a small improvement towards the fully converged 
value of the gap.

On the other hand, however,
in Fig.~\ref{NiO_gap_vs_iters_down} we present the absolute energy
gap, namely, the energy difference between occupied and unoccupied
states independently of the spin character of the band. 
Notice that both calculations, much before they take a ``path'' to final 
convergence, depart significantly from the original values of the gap.
The reason is that the energy levels for up and down spins
are free to vary independently in the present QPscGW calculations. As a 
result, at the iteration where the gap deviates from the monotonically
increasing behavior, we find that the down-spin energy eigenvalue which 
corresponds to the valence band at the X point ($k=(1/2,0,1/2)$) 
starts rising, thus, entering ``inside''
the gap leading to level crossing. This is illustrated in 
Fig.~\ref{NiO-level-crossing}, where we plot the bands at the 43$^{rd}$ 
QPscGW iteration at which the energy of the top-most valence 
band for the down-spin state at the X point rises. 
This is a somewhat similar
behavior to that discussed in the case of CoO.  Eventually, as the
QPscGW iteration process continues, this energy level is lowered again and
the QPscGW procedure converges. 

In Fig.~\ref{NiO_DOS_SCGW} we compare the converged density of states 
obtained from the
QPscGW calculations starting from HSE06 and sGGA.
Notice that the agreement is very good.
As we will see in the following section, the converged solution for
the density of states compares rather well with the experimental 
photoemission results.

%

\subsection{Self-energy and spectral functions}
\label{self}
\begin{figure*}[htp]
\vskip 0.3 in
     \begin{center}
        \subfigure[]{%
            \label{fig:MnO_self}
            \includegraphics[width=\figw]{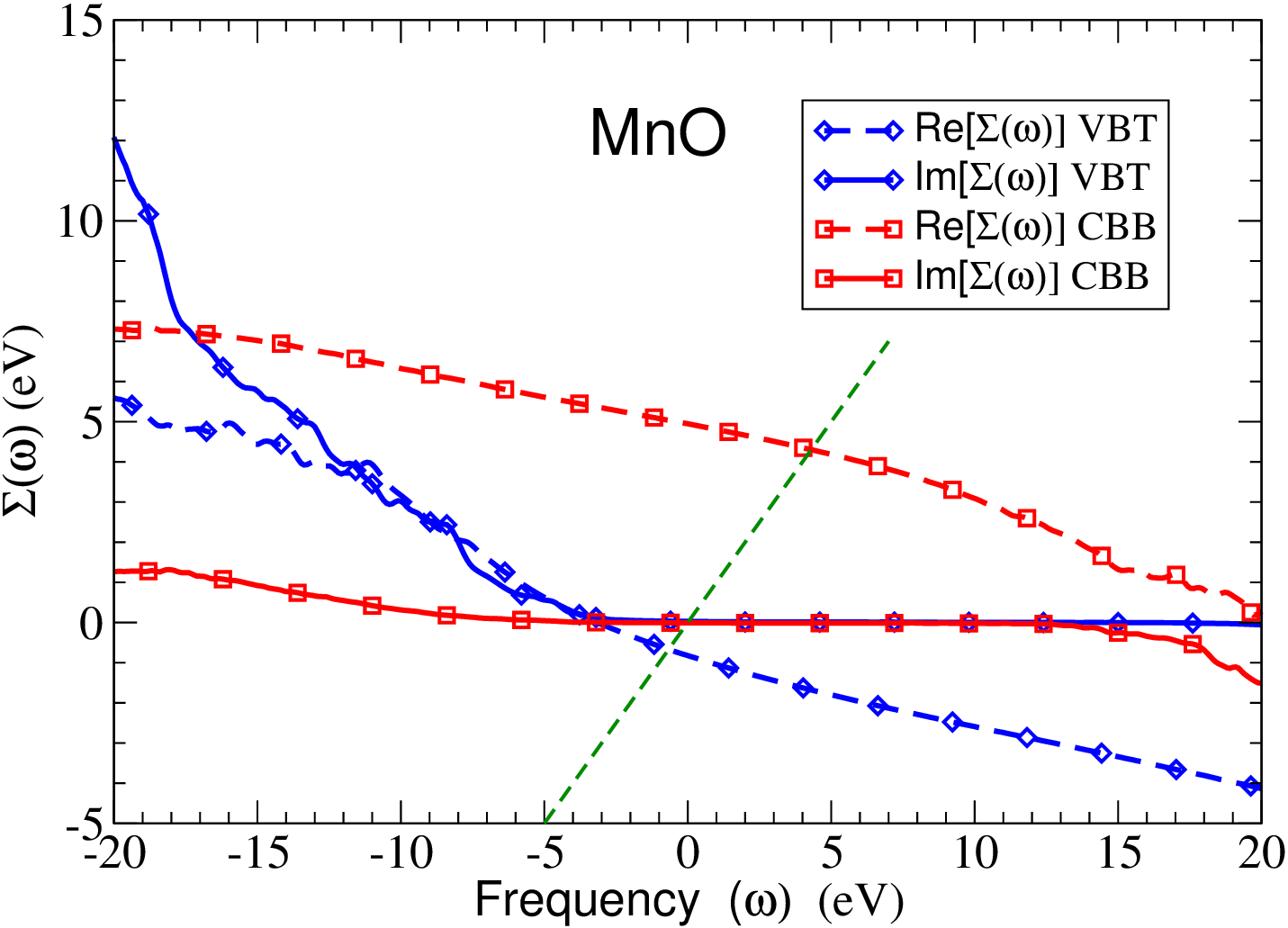}
        }%
\hskip 0.05 in        \subfigure[]{%
            \includegraphics[width=\figw]{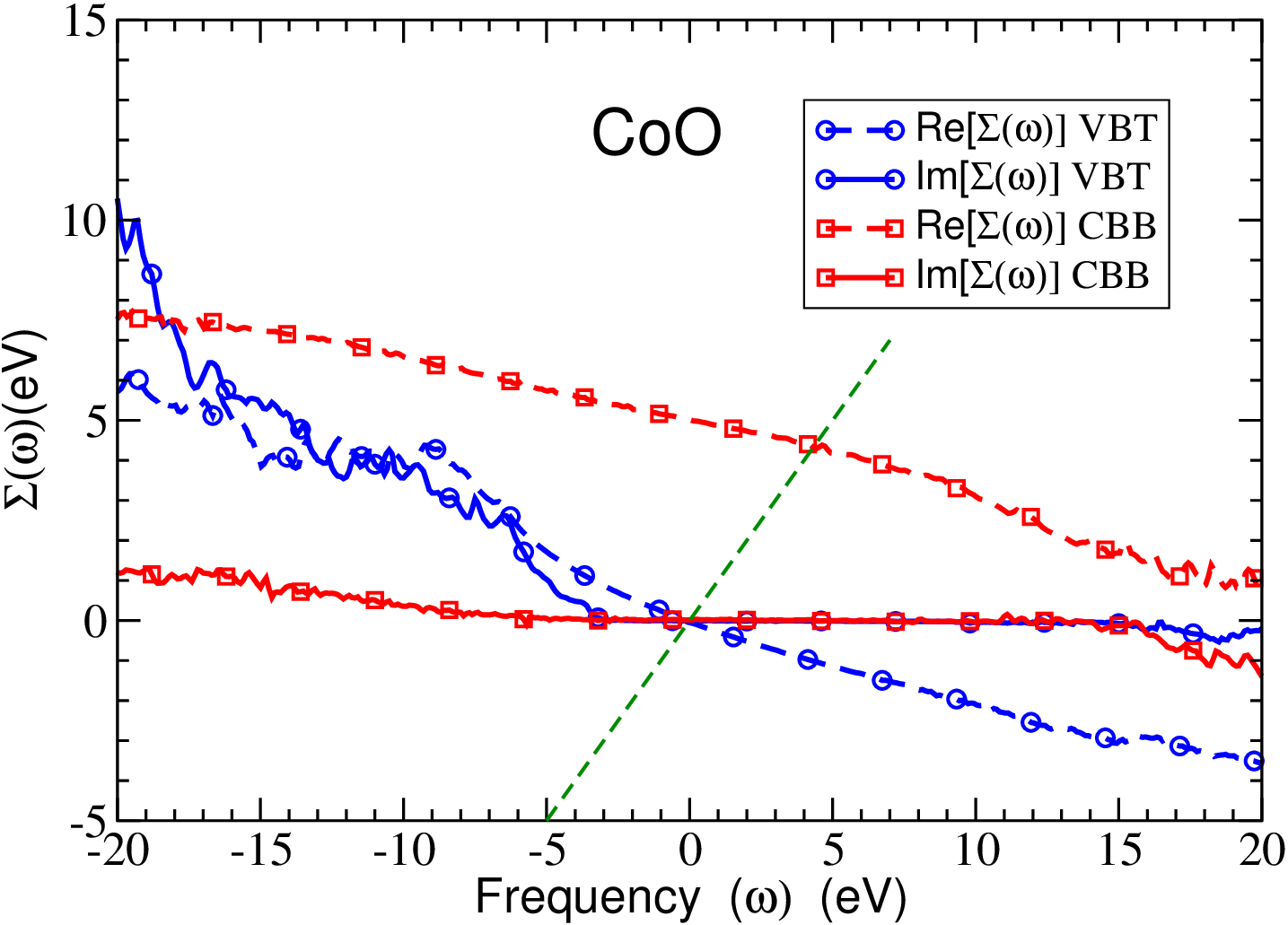}
            \label{fig:CoO_self}
        }%
\hskip 0.05 in        \subfigure[]{%
            \includegraphics[width=\figw]{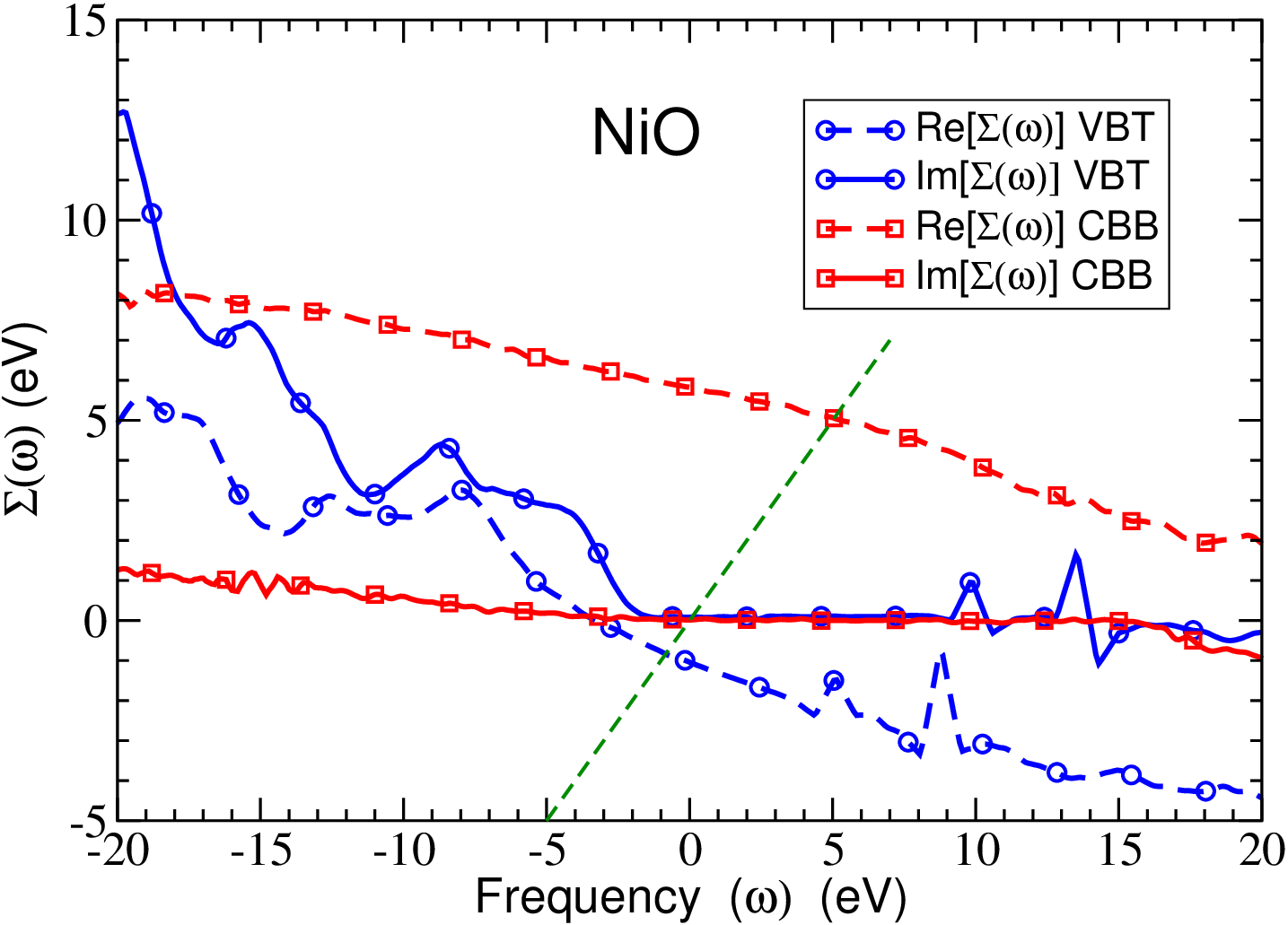}
            \label{fig:NiO_self}
        }   \\
\vskip 0.3 in
        \subfigure[]{%
            \label{fig:MnO_G}
            \includegraphics[width=\figw]{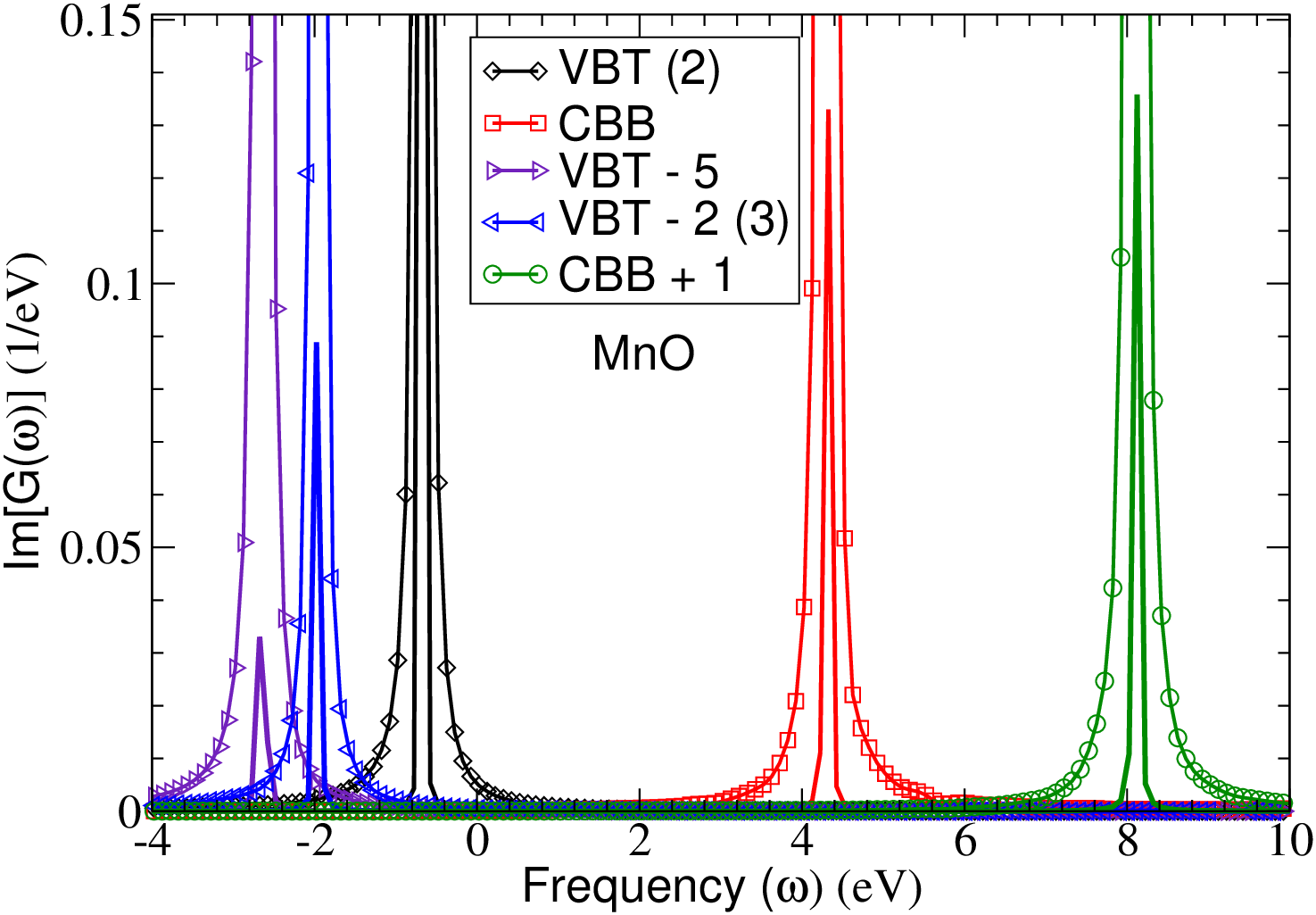}
        }%
 \hskip 0.05 in        \subfigure[]{%
            \includegraphics[width=\figw]{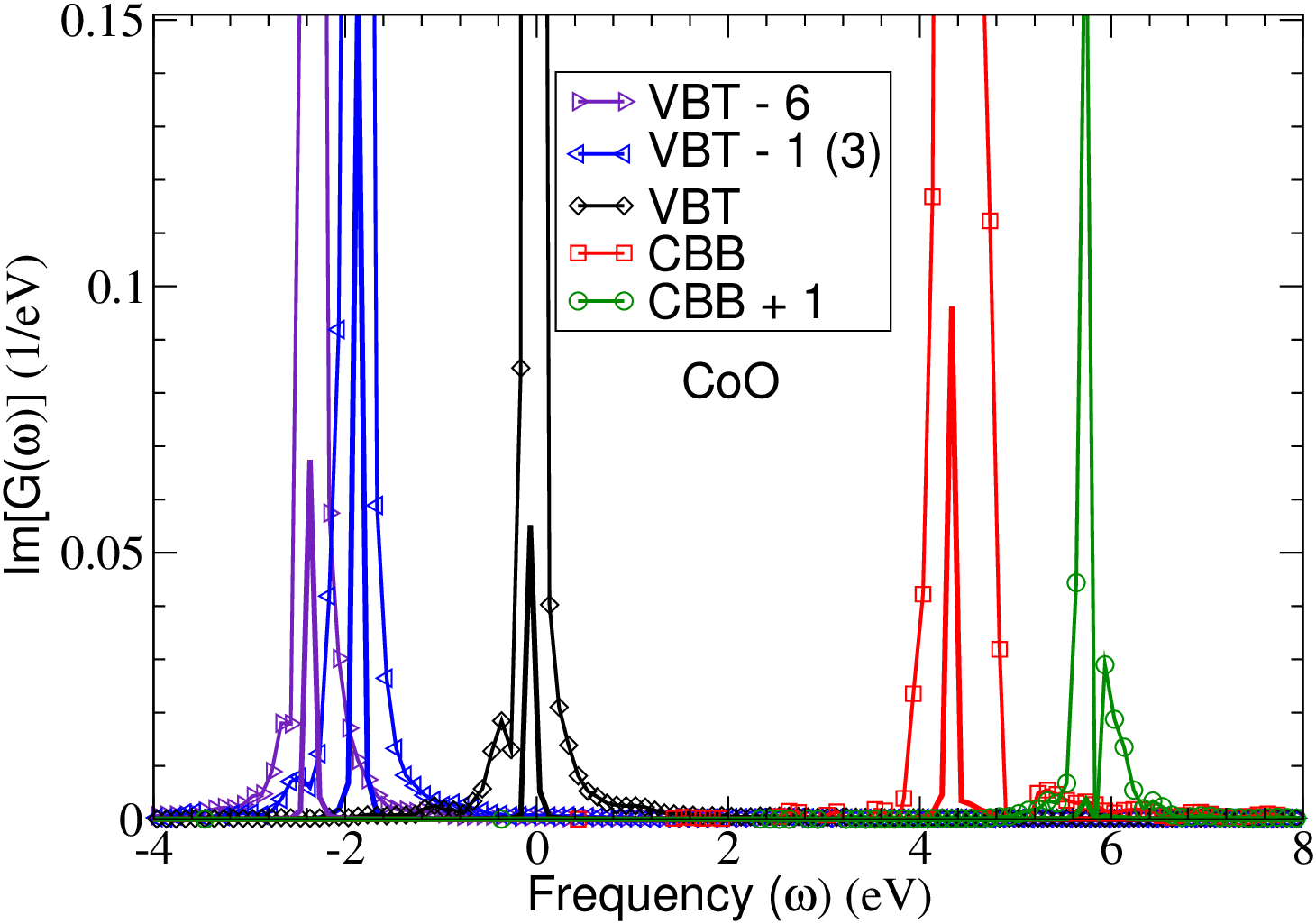}
            \label{fig:CoO_G}
        }%
 \hskip 0.05 in        \subfigure[]{%
            \includegraphics[width=\figw]{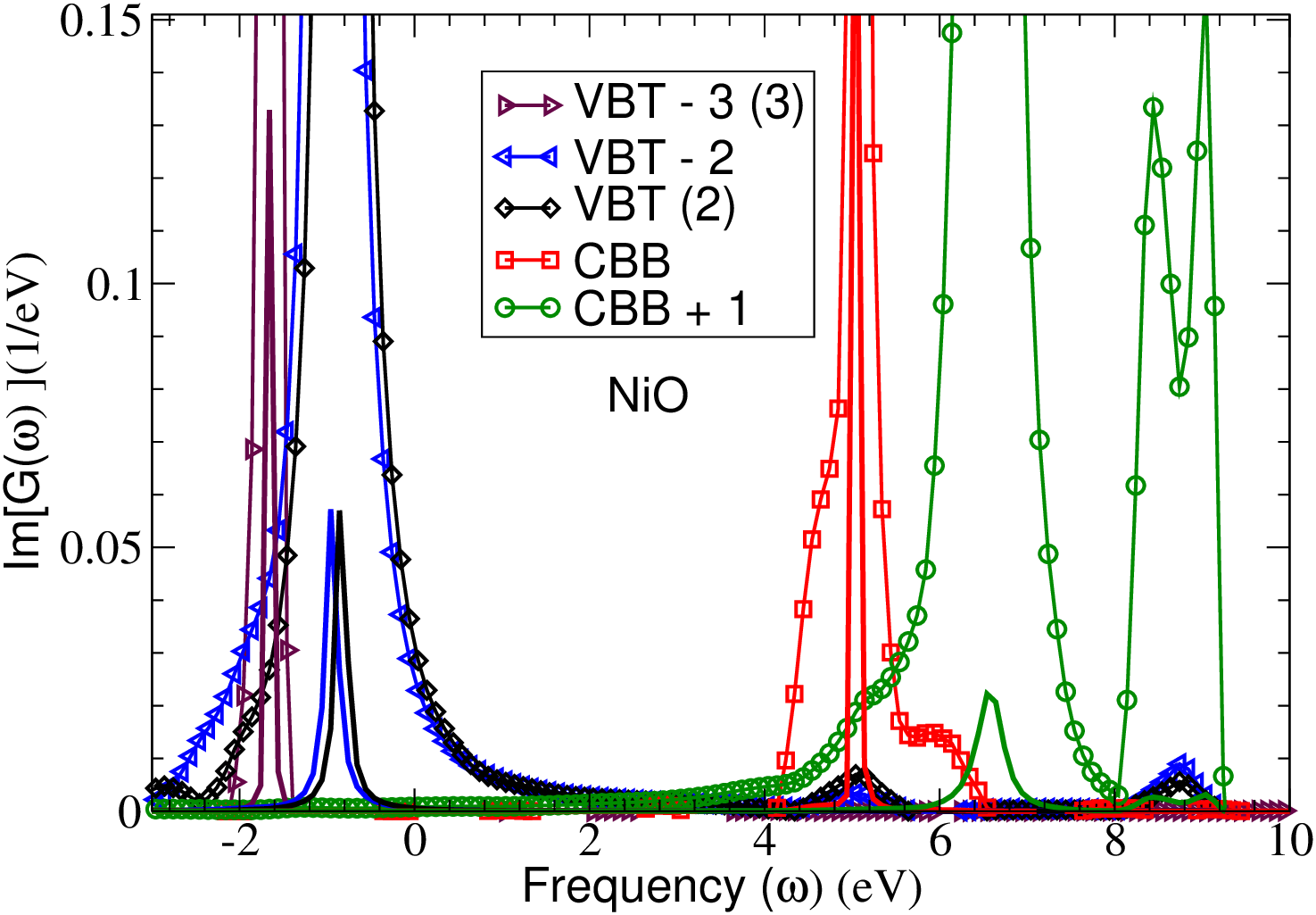}
            \label{fig:NiO_G}
        }%
    \end{center}
\caption{(color-online)Top: The real and imaginary parts of
the self-energy are shown for the three TMOs for the topmost valence band (VBT)
and the lowest conduction band (CBB) at the $\Gamma$ point. 
Bottom: The imaginary part of
$G$ for the VBT (circles joined by blue lines) and 
for the CBB (circles joined by blue lines) at the $\Gamma$ point. In addition, 
the imaginary part of $G$  for a few other bands are shown. 
These curves are labeled as CBB+n or VBT-n, where $n$ denotes
the order of the band above or below the lowest conduction or highest 
valence band respectively. The degeneracy of each state is given in
 parentheses if it is different from 1. Under each curve the same curves are
plotted in plain solid lines and scaled down by a factor of 50.}
\label{all-self}
\vskip 0.3 in
\end{figure*}

In Fig.~\ref{all-self}(a-c) we present the calculated real and imaginary parts
of the self-energy at the $\Gamma$ point for
the highest valence band (circles joined by blue lines)  and 
the lowest conduction band 
(square joined by red lines). 
The intersection between the  the 45$^{\circ}$ sloped green dashed-line 
and the $Re\Sigma$ yields the value of the on-shell value of the 
real-part of $\Sigma$. Notice that,  the imaginary part of $\Sigma$ is small
for values of $\omega$ near the quasiparticle peaks.
In Figs.~\ref{fig:MnO_G},\ref{fig:CoO_G},\ref{fig:NiO_G}
 the imaginary part of the Green's function  at the $\Gamma$ point is
presented for the lowest conduction band (squares joined by red lines)
and the topmost valence band (circles joined by blue lines).
In addition,  the imaginary part of $G$ is plotted for the same
bands that are shown in Fig.~\ref{band_types}. 
All curves scaled down by a factor
of 50 are also plotted. Notice that the half-width at the quasiparticle
peaks is small and it is the imaginary part of the self-energy at
the quasiparticle peak. This indicates that a) the quasiparticle
states are relatively well-defined near the QP peaks, and b) neglecting
the imaginary part of $\Sigma$ in Eq.~\ref{hamiltonian}, and 
Eq.~\ref{exc-corr}, may be a justified approximation for these materials.

\section{Fully Converged QPscGW Results}
\label{results}

\subsection{Bands, gaps and magnetic moments}
In Fig.~\ref{all-bands}, the bands for MnO,  CoO, and NiO
are shown, as obtained from sGGA or GGA+U (solid lines)
and from QPscGW (up and down triangles). The bands slightly break  
the up-down symmetry as discussed in Section~\ref{convergence}.

In MnO, CoO and NiO, the conduction band minimum(CBM) at the $\Gamma$ 
point is found to be of purely $s$ character. 
In the conduction band, as we move away from the $\Gamma$ point, we 
find a significant contribution from the Mn $t_{2g}$ states and 
Ni $e_g$ states in the case of MnO and NiO respectively. For CoO, the 
character of the conduction band changes from $s$ to Co $t_{2g}$  
as we move away from 
the $\Gamma$ point.

In the case of MnO and NiO, the valence band maximum (VBM) is dominated by 
transition metal $e_g$ states along with some mixture of O $p$ states. 
In CoO, the VBM has mostly  Co $t_{2g}$ character with 
considerable admixture of $e_g$ states and hybridized with O $p$.
As is apparent from  Fig.~\ref{all-bands}, the bands near the
Fermi energy of all three materials are 
very flat, which makes it difficult to determine the exact nature of 
the VBM within the accuracy of the QPscGW calculation. 

In Fig.~\ref{fig:first-types} the band order for MnO at the $\Gamma$ point
is given for sGGA, and for GGA+U with various values of
U, along with the results of our QPscGW calculation starting from sGGA.
Notice that as a function of U, the GGA+U calculation leads to the same
band-type ordering as sGGA for states near the Fermi level, except for large
values of U (7 eV) where another $s$-type band with some O $p$ admixture 
comes into
play. The same $s-p$ band appears in the converged state of the QPscGW
calculation as shown in Fig.~\ref{fig:first-types}.
Since this behavior appears in the GGA+U calculation for large U, 
it can be understood in the following way. At such high values of U, in
the GGA+U calculation the large on-site Coulomb repulsion separates
the energy eigenvalues of the  $t_{2g}$ from 
the $e_{g}$ states, such that the O $p$ state
surfaces and mixes with a significant portion of the Mn $e_g$ state
to form the top valence band. For the same reason, the conduction band
of $s-p$ character comes down because the $t_{2g}$ states are pushed up.
For significantly greater values of U (15 eV) 
the topmost valence band becomes 
a pure O $p$ band and the $t_{2g}$ conduction band is pushed even higher. 
Therefore, we conclude that a simple  GGA+U calculation qualitatively describes the 
orbital content of the bands near the Fermi level in MnO at the $\Gamma$ point.
We should stress, however, that in this regime, where the orbital character 
between QPscGW and GGA+U match, the band-gaps do not match in size.
The value of the gap obtained for GGA+U with U=7 eV is 2.35 eV and the 
gap obtained with  QPscGW is $\sim$~4.4 eV. However, if one forces GGA+U to describe the correct
size of the gap by choosing an approximate value of U, 
then one gets into a regime with the wrong orbital content near the
Fermi energy.

The nature of the lowest conduction bands and the highest occupied bands
for the case of CoO  has been discussed thoroughly in 
Sec.~\ref{convergence}. The reader is referred to that section for the 
illustration of the interesting physics arising from the
hybridization and the interplay between the lowest conduction band,
which is of Co $t_{2g}$ character and is almost dispersionless, 
and the next lowest but much more dispersive conduction
band, which is of Co $s$ character. This hybridization gives rise to
dispersion in the vicinity of the $\Gamma$ point.
As mentioned in  Sec.~\ref{convergence}, this hybridization occurs
as a function of U in the GGA+U approximation 
for values of U $>$ U$_c$ (where U$_C$ is between 3 and 4 eV). 
The physics of the actual material,
CoO, resides in the regime of U  $>$ U$_c$.
Again, we want to stress that, while the physics of CoO as
seen by the fully convergent QPscGW calculation and the GGA+U 
is similar, in the regime of U where this happens, 
the size of the GGA+U band gap
is significantly smaller than the QPscGW gap. 

Fig.~\ref{fig:fourth-types} illustrates that the
level ordering of NiO, as obtained by the fully converged
QPscGW calculation starting from either GGA or HSE06, is
the same as that obtained by the HSE06 calculation.
The conduction band ordering is also similar to 
that obtained by the GGA+U calculation. The nature of the
highest valence bands obtained by GGA+U for U = 4 eV is similar to 
the QPscGW results for two of the three topmost Ni valence bands 
of $d$ character at $\Gamma$.
If we increase the value of U to 8 eV, the nature of the topmost
valence bands becomes O $p$ type at the $\Gamma$ point, 
which is qualitatively different  from the character 
of these bands obtained from the QPscGW 
calculation. The reason for this may be the fact that a large
value of U pushes the Ni $d$ states away from each other,
which brings the O $p$ states to the surface because
they are not affected by the large value of U.

\begin{table}[htp]
\begin{tabular}{|p{70pt}|p{50pt}|p{50pt}|p{50pt}|} \hline
 \textbf{Gap (eV)}    & \textbf{MnO}  & \textbf{CoO}  & \textbf{NiO} \\ \hline
         
 GGA (indirect)                & 0.9    & 0.0   & 0.7  \\ \hline
 GGA+U (U=4 eV) (indirect)                &    & 2.81   & ---  \\ \hline
 HSE03 (indirect) \cite{Rodl-G0W0+HSE}               & 2.6   & 3.2  &  4.1  \\ \hline
 HSE03 (direct) \cite{Rodl-G0W0+HSE}               &   3.2   & 4.0  &  4.5  \\ \hline
 HSE03+G0W0 (indirect) \cite{Rodl-G0W0+HSE}          & 3.4   & 3.4   & 4.7  \\ \hline
 HSE03+G0W0 (direct) \cite{Rodl-G0W0+HSE}          & 4.0   & 4.5   & 5.2  \\ \hline
 QSGW (indirect) \cite{Schilfgaarde}                 & 3.5    & ---  &  4.8  \\ \hline 
 mGW (indirect) \cite{Asahi}                 & 4.03  & 3.02  & 3.6  \\ \hline
 LDA+U+GW (indirect) \cite{Rinke}            & 2.34  & 2.47  & 3.75  \\ \hline
 \textbf{Present QPscGW (indirect)} & 4.39   & 4.78$\pm$0.13    & 5.0   \\ \hline
 Photoemission        & 3.9$\pm$0.4\cite{Berger}  & 2.5$\pm$0.3\cite{Elp}  & 4.3\cite{Allen} \\ \hline
 Conductivity         & 4.0$\pm$0.2\cite{Drabkin}  & 3.6$\pm$0.5\cite{Gvishi} &
  3.7\cite{Ksendzov} \\ \hline
 Optical Absorption   & 3.7$\pm$0.1\cite{Ksendzov}   & 2.8\cite{Pratt}, 5.43\cite{Kang}  & 3.7\cite{Powell}, 3.87\cite{Kang} \\ \hline

\end{tabular}
\caption{Comparison of the energy band gaps as obtained in the present work with other methods along with various experimental results. }
\label{table1}
\end{table}

\begin{table}[htp]
\begin{tabular}{|p{70pt}|p{50pt}|p{50pt}|p{50pt}|} \hline
 \textbf{Moment ($\mu _{B}$)}      & \textbf{MnO}  & \textbf{CoO}  &  \textbf{NiO} \\ \hline
         
 GGA                  & 4.34    & 2.4  & 1.3  \\ \hline
 GGA+U  (U=4 eV)              &    & 2.73   &   \\ \hline
 HSE03\cite{Rodl-G0W0+HSE}                & 4.5    & 2.7   & 1.6  \\ \hline
 QSGW\cite{Schilfgaarde}                  & 4.8   & ---   & 1.7  \\ \hline 
 mGW\cite{Asahi}                  & 4.56  & 2.61 &  1.57  \\ \hline
 \textbf{Present (QPscGW)} & 4.58   & 2.74   & 1.7   \\ \hline
 Experiment           & 4.58\cite{MnOlattice}  & 3.35\cite{Khan}, 3.8\cite{Roth}$^,$\cite{Burlet}, 3.98\cite{CoOlattice}  & 1.9\cite{MnOlattice}$^,$\cite{Roth} \\ \hline

\end{tabular}
\caption{Comparison of the magnetic moments as obtained in the present 
work with other methods along with various experimental results. }
\label{table2}
\end{table}

In Table~\ref{table1}, the band gaps obtained from our QPscGW 
approach are compared with various other parameter-dependent and 
\emph{ab initio} methods. For MnO, we started with a gap of 0.9~eV 
obtained from an sGGA calculation and the converged QPscGW 
calculation yields a gap of 4.39~eV, which 
is somewhat larger than but in reasonably good agreement with 
the observed photoemission results within 
experimental error. For NiO, the band gap obtained by our calculations 
is 5.0~eV, which is somewhat larger than the experimental value of 4.3~eV. 
For CoO, our QPscGW calculations starting with wavefunctions and 
quasiparticle energies  obtained from a GGA+U with U = 2, 3, 4 or 6 eV, 
yield a band gap of 4.78$\pm0.13$~eV. This band-gap also overestimates the 
experimental gap. 
Therefore, we find a systematic overestimation of the band-gaps of 
these TMOs.
This is expected due to the missing vertex corrections\cite{Shishkin3}
and reduced quasiparticle screening. 

We would also like to note that our results concerning the band
character and ordering are in excellent agreement with those from
LDA$^{\prime}$+DMFT\cite{LDA'+DMFT}, as can be inferred from the 
projected density of
states and band structures given in the work cited above.

In Table~\ref{table2}, the local magnetic moments of MnO, CoO and NiO
are compared with their values obtained using different approaches. 
The calculated magnetic moment for MnO using 
the QPscGW method  is in excellent agreement with experimental value. 
The value of the moment obtained for CoO is significantly lower than the 
experimental value and this may be attributed to the orbital 
contribution to the magnetic moment.\cite{Rodl-G0W0+HSE}
Calculations at the GGA+U level including the
orbital contribution significantly increases the moment to 3.58, which is 
within the range of experimentally obtained values.

\subsection{Density of states}
Fig.~\ref{dos_vs_expt} shows the self-consistently converged density of 
states for these TMOs as compared to experimental photoemission 
experiments. The center of the gap of the experimental photoemission data 
is aligned with that of the calculated density of states. 
The dot-dashed lines are the calculated DOS where a Gaussian 
broadening with 0.4 eV full-width has been applied. 
The solid lines
correspond to the calculated DOS where a Gaussian 
broadening with the experimental resolution
has been applied. The experimental resolution depends on both material
and XPS vs BIS measurements. These are taken directly from the corresponding 
references for each of the materials\cite{Berger,Elp,Allen}. In addition, we have
scaled the intensities of the XPS and BIS separately to approximately mimic the amplitudes
of the DOS. Notice that the experimental resolution for both XPS and BIS is poor ($\sim 1 eV$) and,
thus, determining the size of the gaps (given in Table~\ref{table1}) and the location of the Fermi level
is difficult. Thus, instead of aligning the experimental and calculated Fermi levels
we chose to align the center of the gaps. 
The density of states is in reasonable overall agreement with the intensity of the experimental 
XPS and BIS data. The QPscGW calculation leads to an  overestimation of 
band-gaps due to the neglect of 
the vertex corrections in the GW 
approximation\cite{Shishkin3,Schilfgaarde}.

\section{CONCLUSIONS AND SUMMARY}
\label{conclusions}
We have studied the electronic structure of the $d$ electron systems MnO, 
 CoO, and NiO by means of the QPscGW approach with the aim to answer the
convergence questions outlined in the Introduction of this paper. 

First, we studied the convergence of the QPscGW procedure starting from GGA+U
with different values of U. Using a rather wide range of values of
U (2-6 eV) for CoO, we find that our converged results for
bands and density of states are weakly dependent on the starting 
GGA+U solutions. The fully converged energy bands and density of states,
starting from different values of U, are very close to each other. 
In addition, we found that starting our QPscGW procedure from  GGA+U
wavefunctions and eigenvalues can improve the convergence if a
suitable value of U is chosen.

We also studied the convergence of QPscGW for NiO
starting from  HSE06 quasiparticle states and from quasiparticles
as obtained from an sGGA calculation. We find that
the results for the bands and the density of states are in 
very good agreement with each other.
Interestingly, we find that during the QPscGW evolution towards
the converged solution, certain levels cross and this makes the
``path'' temporarily diverge from the approach to the converged solution.

We find that single-shot G$_0$W$_0$ calculations are somewhat
insufficient to bridge the difference between the gap obtained within 
any of the starting states (including when we start from HSE) 
and the converged QPscGW gap. Namely, even a G$_0$W$_0$ on top  of
HSE gives only a fraction of the correction needed to bridge 
the {\it difference} between the gap obtained at the HSE level 
and the gap obtained at the fully converged level. 

This convergence
study as a function of U also demonstrates that the physics
of CoO for the bands near the Fermi level is very similar to
that obtained from a GGA+U calculation for values of U above
4 eV where a hybridization occurs between a lower energy dispersionless
conduction band of Co $t_{2g}$ character and a much more dispersive 
Co $s$ band.

For the case of MnO, we find that within a simple GGA+U treatment,
there is a region of U where the orbital character of the bands near
the Fermi level and their relative ordering, as obtained by the fully converged QPscGW 
approach, can be reproduced.
For NiO, the band ordering near the Fermi level 
obtained by the QPscGW calculation can only be partially described by
means of a simple GGA+U calculation using a U only on the $d$ levels. 

The magnetic moments obtained by solving the GW equations 
self-consistently agree reasonably well with experimental results. 
The calculated band-gaps are somewhat overestimated
as expected\cite{Shishkin3,Faleev}
 due to reduced quasiparticle screening and neglect of vertex corrections. 
Furthermore, the calculated density of states, determined from the 
converged wavefunctions and quasiparticle energies, agrees reasonably 
well with the results of photoemission experiments.  We have also shown 
that the self-consistently determined wavefunctions and energy 
gaps are weakly dependent on the starting wavefunctions. 

We conclude that this  approach to transition metal oxides with 
$d$ states at the Fermi level may be computationally 
demanding, but it is a genuinely parameter-free approach and provides a good 
prediction for energy gaps, magnetic moments, density of states, and 
quasiparticle wavefunctions. 
In addition, such an approach is quite important for 
testing the simpler picture suggested from GGA+U calculations and can provide
useful input for model studies aiming at describing the low energy physics in
these materials.

We believe that an improvement of the approach could be made by 
generalizing  the approach 
suggested
in Ref.~\onlinecite{Shishkin3} and Ref.~\onlinecite{Faleev} to include 
 vertex corrections for spin-polarized systems.

\section{Acknowledgments}
This work was supported in part by the US National High
Magnetic Field Laboratory, which is partially funded by the
US National Science Foundation.

\end{document}